\documentclass[lettersize,journal]{IEEEtran}
\usepackage{amsmath,amssymb,amsfonts}
\usepackage{algorithmic}
\usepackage{algorithm}
\usepackage{array}
\usepackage[caption=false,font=footnotesize,labelfont=rm,textfont=rm]{subfig}
\usepackage{textcomp}
\usepackage{stfloats}
\usepackage{url}
\usepackage{verbatim}
\usepackage{graphicx}
\usepackage{cite}
\usepackage{xcolor}
\usepackage{subfig}
\def\BibTeX{{\rm B\kern-.05em{\sc i\kern-.025em b}\kern-.08em
    T\kern-.1667em\lower.7ex\hbox{E}\kern-.125emX}}
\usepackage{soul}
\usepackage{booktabs}

\DeclareSubrefFormat{myparens}{#1(#2)}
\begin{document}

\color{black}

\title{Scavenger+: Revisiting Space-Time Tradeoffs in Key-Value Separated LSM-trees}

\author{
    \IEEEauthorblockN{Jianshun Zhang, \textit{Student Member, IEEE,} Fang Wang, \textit{Member, IEEE,} Jiaxin Ou, Yi Wang,}
    
    \IEEEauthorblockN{Ming Zhao, Sheng Qiu, Junxun Huang, Baoquan Li, Peng Fang*, Dan Feng, \textit{Fellow, IEEE.}}
    
\thanks{*Corresponding author: Peng Fang (fangpeng@hust.edu.cn). This work was funded by the National Key R\&D Program of China (No.2023YFB4502801), NSFC (No.U22A2027 and 62402187), China Postdoctoral Science Foundation (No.GZB20240243 and 2024M751009), and Postdoctoral Project of Hubei Province (No.2024HBBHCXA024). }
\thanks{Jianshun Zhang, Fang Wang, Junxun Huang, Baoquan Li, Peng Fang and Dan Feng are with the Wuhan National Laboratory for Optoelectronic, Key Laboratory of Information Storage System, Engineering Research Center of data storage systems and Technology, Ministry of Education of China, School of Computer Science and Technology, Huazhong University of Science and Technology, Wuhan, Hubei 430074, China. Fang Wang is also with Shenzhen Huazhong University of Science and Technology Research Institute, China. (E-mail: \{shunzi, wangfang, junix, bqli, fangpeng, dfeng\}@hust.edu.cn)}
\thanks{Jiaxin Ou, Yi Wang, Ming Zhao and Sheng Qiu are with the ByteDance. (E-mail: \{oujiaxin, wangyi.ywq, zhaoming.274, sheng.qiu\}@bytedance.com)}}
\maketitle

\begin{abstract}
Key-Value Stores (KVS) based on log-structured merge-trees (LSM-trees) are widely used in storage systems but face significant challenges, such as high write amplification caused by compaction. KV-separated LSM-trees address write amplification but introduce significant space amplification, a critical concern in cost-sensitive scenarios. Garbage collection (GC) can reduce space amplification, but existing strategies are often inefficient and fail to account for workload characteristics. Moreover, current key-value (KV) separated LSM-trees overlook the space amplification caused by the index LSM-tree. In this paper, we systematically analyze the sources of space amplification in KV-separated LSM-trees and propose Scavenger+, which achieves a better performance-space trade-off. Scavenger+ introduces (1) an I/O-efficient garbage collection scheme to reduce I/O overhead, (2) a space-aware compaction strategy based on compensated size to mitigate index-induced space amplification, and (3) a dynamic GC scheduler that adapts to system load to make better use of CPU and storage resources. Extensive experiments demonstrate that Scavenger+ significantly improves write performance and reduces space amplification compared to state-of-the-art KV-separated LSM-trees, including BlobDB, Titan, and TerarkDB.
\end{abstract}


\section{Introduction}
Key-Value Store (KVS) is a simple yet efficient data storage model that enables rapid access to corresponding values through keys. It finds extensive applications in numerous domains such as social networks and AI/ML \cite{zhou2020database}. KVS can be categorized into various types. One such type is the hash-based KVS, which employs a hash function to map keys to specific storage locations, offering fast lookup capabilities and making it highly suitable for caching scenarios \cite{hotring}. Alternatively, a B+-tree-based KVS stores all data in leaf nodes, with non-leaf nodes serving solely as indices. This structure enables efficient range queries, making it widely used in traditional database systems \cite{MySQLManual}. However, when handling high-frequency write workloads involving large-scale data, LSM-trees demonstrate unique advantages. Exploiting the preference of storage devices for sequential over random writing \cite{arpaci2018operating}, LSM-trees use in-memory buffers to convert randomly distributed user writes into sequential ones \cite{o1996log}. Due to their superior write performance and low space cost, LSM-trees are widely adopted in data-sensitive applications like distributed databases \cite{cao2020characterizing, chen2022bytehtap} and file system metadata management \cite{aghayev2019file}.

Unfortunately, LSM-trees incur high compaction costs. The compaction process in LSM-trees is essential for maintaining the multi-level structure and data order, requiring frequent disk I/Os. This leads to substantial read and write amplification, attracting significant criticism \cite{lu2017wisckey, raju2017pebblesdb}. To reduce compaction overhead, Key-value separation, as proposed by WiscKey \cite{lu2017wisckey}, is widely used in industrial solutions like BlobDB\cite{blobdb}, Titan\cite{titan}, and TerakDB\cite{terarkdb}. KV-separated LSM-trees reduce the size of the original LSM-tree by storing values in a separate file while retaining only index pairs within the LSM-tree. This approach significantly increases the storage overhead, potentially making it unaffordable in cloud computing environments \cite{zhang2022sa}. Prior research \cite{dong2017optimizing, dong2021evolution} shows that the optimization goal of vanilla LSM-trees has shifted from maximizing performance to improving resource efficiency. This shift in prioritization stems from the understanding that reducing space amplification allows multiple instances to share the same hardware, thereby optimizing SSD bandwidth utilization \cite{dong2017optimizing}. While RocksDB efficiently mitigates space amplification, its write performance lags significantly behind that of KV-separated LSM-trees. For instance, under a workload with fixed-length values of 16KB (Fixed-16K), RocksDB's performance is only 0.18x that of BlobDB \cite{blobdb}. Although KV-separated LSM-trees offer higher performance, they cannot reclaim space through compaction alone, unlike vanilla LSM-trees. They also require coordination with garbage collection (GC), and once the values are separated, compaction becomes oblivious to actual space usage. To optimize performance while effectively managing storage space costs, we propose a solution that strikes a balance between storage cost and write performance.

GC operations in KV-separated LSM-trees reclaim storage space, but incur significant I/O overhead and compete with user requests \cite{chan2018hashkv}. Recent studies have focused on optimizing GC operations for improved foreground performance \cite{chan2018hashkv, li2021differentiated}. However, such efforts frequently overlook space reclamation efficiency, exacerbating space amplification from delayed reclamation \cite{blobdb, li2021differentiated}. For instance, BlobDB and DiffKV employ a compaction-triggered GC/merge policy to mitigate the overhead of GC lookups and write indexes \cite{blobdb, li2021differentiated}. However, this approach results in significant space amplification, as value files must wait for compaction before being reclaimed. Consequently, under a Fixed-4K workload, BlobDB leads to 3.4× space amplification, which is significantly higher than in other KV-separated LSM-trees. Additionally, existing research overlooks another factor contributing to space amplification in KV-separated LSM-trees: index LSM-tree space amplification, which determines the extent of hidden garbage accumulation. Existing solutions fail to detect this hidden garbage until the corresponding keys are merged in the index LSM-tree, delaying GC scheduling and space reclamation, ultimately causing severe space amplification. For instance, under a Fixed-8K workload, the space amplification of the index LSM-tree constitutes 48.8\% of the total space amplification. Furthermore, workload variations pose challenges to efficient GC operations and rapid space reclamation. In summary, significant I/O overhead in GC, substantial space amplification in the index LSM-tree, and workload adaptability pose significant challenges to balancing space and performance tradeoffs.

To tackle these challenges, we introduce Scavenger+, offering better space-time trade-offs for KV-separated LSM-trees. Unlike previous approaches, we are the first to systematically investigate the space amplification problem in KV-separated LSM-trees. First, we conduct a systematic analysis of the root causes of space amplification in KV-separated LSM-trees. Then, we minimize space overhead by enhancing GC efficiency without degrading performance. To this end, we optimize both GC operations and the compaction strategy to accelerate space reclamation while preserving foreground performance. This paper extends our previous work \cite{zhang2024scavenger} in three main areas: (1) We consider medium-sized value workloads and propose an adaptive readahead for GC to further reduce the I/O overhead. (2) We fully utilize available CPU and I/O resources and propose a dynamic GC scheduling strategy to accelerate space reclamation without compromising performance. (3) We expand our evaluation to investigate the impact of our approach on tail latency and the individual contribution of each design component. Our main contributions are summarized as follows:
\begin{itemize}
    \item We model and quantify two sources of space amplification in KV-separated LSM-trees, namely exposed garbage in value data and space amplification in the index LSM-tree, thereby identifying two critical operations that impact space amplification: GC and compaction.
\end{itemize}
\begin{itemize}
    \item We propose an I/O-efficient GC scheme that accelerates three critical steps: read, lookup, and write. It minimizes I/O overhead, accelerates GC execution, and facilitates space reclamation, thereby reducing space amplification.
\end{itemize}
\begin{itemize}
    \item We propose a space-aware compaction strategy based on a notion of compensated size. This strategy integrates value storage back into the LSM-tree, allowing standard compaction to effectively control space amplification in the index LSM-tree while improving GC efficiency.
\end{itemize}
\begin{itemize}
    \item We dynamically schedule compaction and GC based on the space amplification impact experienced by the index LSM-tree and value data, respectively. Simultaneously, we throttle GC bandwidth to reduce contention with user requests, thereby utilizing available resources to achieve better trade-offs between performance and storage cost.
\end{itemize}

This paper is organized as follows. Section \uppercase\expandafter{\romannumeral2} analyzes the motivation and quantifies the sources of space amplification. Section \uppercase\expandafter{\romannumeral3} presents the design of Scavenger+. Section \uppercase\expandafter{\romannumeral4} presents the evaluation results. Section \uppercase\expandafter{\romannumeral5} discusses related work. Finally, Section \uppercase\expandafter{\romannumeral6} concludes the paper.

\section{PRELIMINARIES}
\subsection{Log-Structured Merge-tree}

\begin{figure*}[tbp]
\begin{minipage}[b]{.64\textwidth}
    \setlength{\abovecaptionskip}{5pt}
    \setlength{\belowcaptionskip}{-0.8cm}
    \centering
    \subfloat[RocksDB]{
            \includegraphics[width=.3\textwidth]{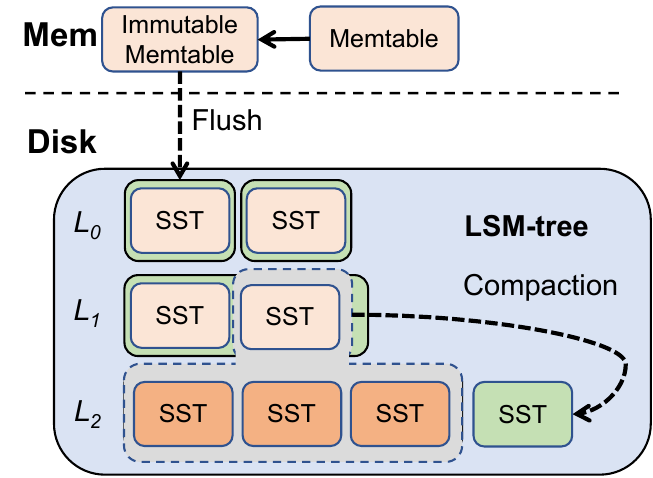}
            \label{fig:bg1-lsm}
        }
    \subfloat[WiscKey]{
            \includegraphics[width=.3\textwidth]{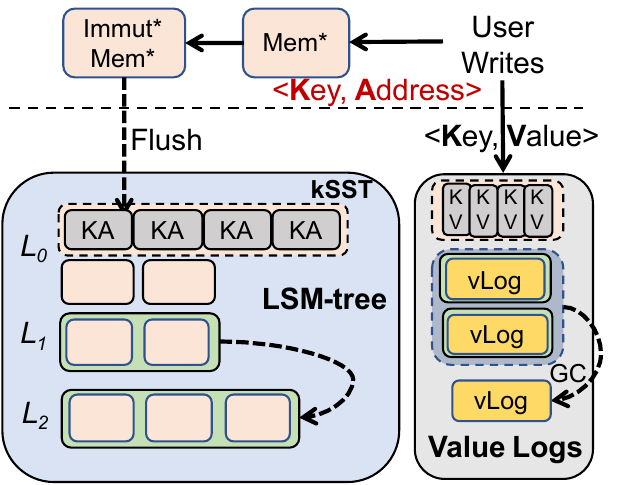}
            \label{fig:bg2-wisc}
        }
    \subfloat[TerarkDB]{
            \includegraphics[width=.3\textwidth]{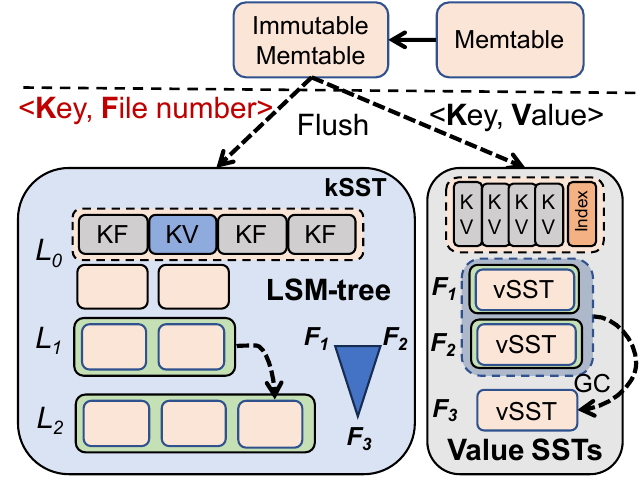}
            \label{fig:bg2-tdb}
        }
    \label{fig:bg1}
    \caption{Overview of RocksDB, WiscKey and TerarkDB}
    \vspace{-0.5cm}
\end{minipage}
\begin{minipage}[b]{.33\textwidth}
    \setlength{\abovecaptionskip}{5pt}
    \setlength{\belowcaptionskip}{-0.8cm}
    \centering
    \includegraphics[width=.9\textwidth]{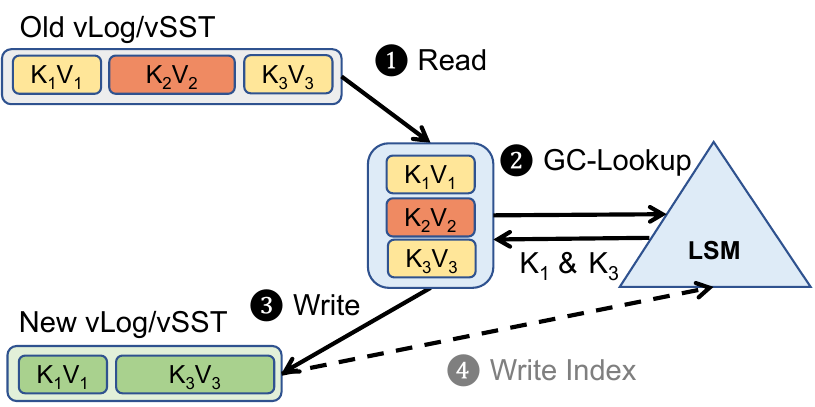}
    \label{bg-gc}
    \caption{Workflow of GC operations.}
    \vspace{-0.5cm}
\end{minipage}
\end{figure*}

Vanilla LSM-trees \cite{o1996log} typically utilize a hierarchical structure consisting of Memtables in memory and Sorted String Tables (SSTables, abbreviated as SSTs) on disk, as illustrated in Figure~\subref*{fig:bg1-lsm}. By leveraging Memtables, LSM-trees can buffer user writes and subsequently flush them to storage as SSTables, thereby transforming random writes, which degrade storage performance, into sequential writes. Concurrently, LSM-trees organize data into an ordered hierarchy from levels $L_1$ to $L_N$ on the storage device, allowing queries to efficiently access level $L_i$, where relevant data resides.

Although LSM-trees provide fast writes and efficient queries, they also introduce considerable I/O overhead. This overhead arises from maintaining the leveled hierarchy of LSM-trees, necessitating frequent background compaction tasks. These tasks involve merging and sorting data across levels, reclaiming space from obsolete data, and preserving data order. However, this process, involving repeated SST reads and writes, significantly increases write amplification. This high write amplification leads to inefficient disk bandwidth utilization for user requests, highlighting a critical opportunity to improve LSM-tree write performance.

\subsection{Key-Value Separation: WiscKey and TerarkDB}
To mitigate write amplification in conventional LSM-trees, WiscKey \cite{lu2017wisckey} introduced key-value separation, which reduces LSM-tree size and minimizes compaction overhead. This technique has been widely adopted in industrial systems such as BlobDB \cite{blobdb}, Titan \cite{titan}, and TerarkDB \cite{terarkdb}. Moreover, numerous studies on LSM-tree optimization have leveraged key-value separation to enhance performance \cite{chan2018hashkv, li2021differentiated}.

\textbf{WiscKey}. As illustrated in Figure~\subref*{fig:bg2-wisc}, WiscKey handles write operations by appending key-value (KV) data to a value log (\textbf{vLog}) file. Subsequently, it stores the corresponding address within the vLog, along with the key as a key-address (\textbf{KA}) entry in the index LSM-tree, formatted as \textit{\(\langle \text{key}, \text{ file\_number}, \text{ offset} \rangle\)}. Read requests first query the index LSM-tree to retrieve the key’s associated address, then access the specified offset within the corresponding vLog file to obtain the value. Garbage collection (GC) in WiscKey involves sequentially scanning the vLog file, extracting key-value entries, and querying the LSM-tree to verify their validity by comparing addresses with the current scanning position. If an entry is valid, it must be rewritten into a new vLog file, and the corresponding address in the index LSM-tree must be updated accordingly. Otherwise, invalid entries can be safely discarded, and once GC is complete, obsolete vLog files can be deleted. Similar to WiscKey, both BlobDB and Titan store addresses in the index LSM-tree. Consequently, their GC operations inevitably necessitate writing back index entries.

\textbf{TerarkDB} is a high-performance key-value store developed by ByteDance, serving as the underlying storage engine for cloud-native databases \cite{chen2022bytehtap} and stream processing systems \cite{flink}. As shown in Figure~\subref*{fig:bg2-tdb}, incoming data is first buffered in an in-memory Memtable. During the flushing process, values are written to ordered value SSTables (\textbf{vSST}), while key-file (\textbf{KF}) indices in the format of \textit{\(\langle \text{key}, \text{ file\_number} \rangle\)} are stored in key SSTables (\textbf{kSST}) of the index LSM-tree. For read operations, the system queries the index LSM-tree to obtain the file number associated with the target key, locates the corresponding vSST, and retrieves the value by resolving the physical location within the vSST through its internal index.

For garbage collection (GC), TerarkDB calculates the global garbage ratio, defined as the proportion of invalid key-value pairs, after each LSM-tree compaction, and triggers GC when the ratio exceeds a predefined threshold (e.g., 20\%). The GC process identifies the vSST file with the highest garbage ratio, scans its key-value pairs, and queries the index LSM-tree to validate each key. If the key’s associated file number in the index matches the current vSST, the entry is considered valid and is rewritten to a new vSST. Instead of updating the index, TerarkDB maintains a file number mapping that redirects the original vSST (e.g., $F_1$ or $F_2$) to the new vSST (e.g., $F_3$). This mapping, illustrated by the triangle in Figure~\subref*{fig:bg2-tdb}, allows read operations to locate the latest valid vSST without modifying the index. By decoupling data reorganization from index maintenance, this design avoids write-back overhead and reduces contention between GC and foreground writes.

\subsection{High Space Amplification of KV-separated LSM-trees}

\begin{figure}[tbp]
    \setlength{\abovecaptionskip}{5pt}
    \setlength{\belowcaptionskip}{-0.6cm}
    \centering
    \subfloat[Update throughput]{
            \includegraphics[width=.47\columnwidth]{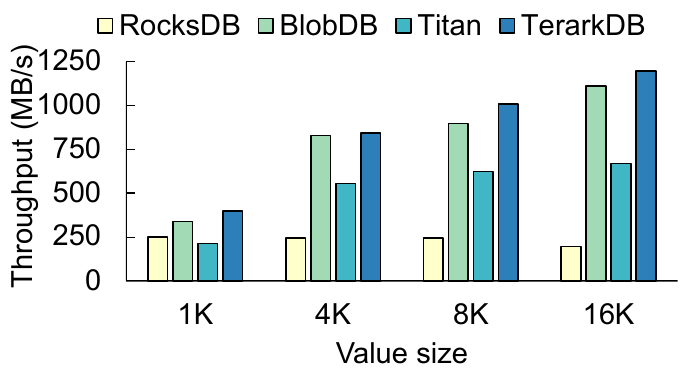}
            \label{fig:mot1-1}
        }
    \subfloat[Space amplification]{
            \includegraphics[width=.47\columnwidth]{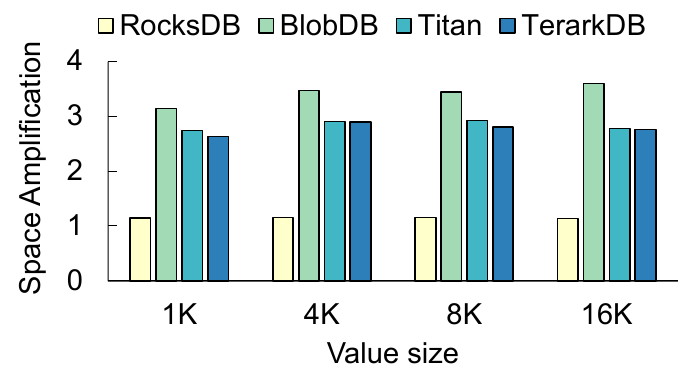}
            \label{fig:mot1-2}
        }
    \caption{Space-time trade-offs of existing solutions.}
    \label{mot1}
    \vspace{-0.5cm}
\end{figure}

While KV-separated LSM-trees minimize the amount of data involved in compaction to reduce overall write amplification and improve foreground write throughput, they lead to increased space overhead. Despite incorporating GC to reclaim storage space, KV-separated LSM-trees still exhibit high space amplification, which significantly exceeds that of the vanilla LSM-tree. To demonstrate the trade-offs between time and space in existing KV-separated LSM-trees, we evaluated the update performance and space amplification of various KV-separated implementations, including BlobDB, Titan, TerarkDB, and the widely adopted LSM-tree, RocksDB. As depicted in Figure \ref{mot1}, KV-separated LSM-trees can significantly enhance write performance, especially for workloads with large values, but at the cost of increased space usage. For a fixed-length 8KB workload, KV-separated LSM-trees can improve update throughput by 2.57× to 4.16× compared to RocksDB. However, this improvement comes at the cost of a 2.42× to 2.97× increase in space overhead. Even with the garbage ratio threshold set to 20\% to trigger GC in KV-separated LSM-trees, which corresponds to an expected space amplification of 1.25 (1/(1-20\%)=1.25), the actual space amplification significantly exceeds this expectation.

KV-separated LSM-trees prioritize write performance over storage efficiency, leading to high total cost of ownership (TCO), challenges in scaling to single-disk multi-instance setups, and suboptimal storage utilization. We aim to balance storage efficiency and performance in KV-separated LSM-trees to accommodate diverse workloads and resource constraints. While KV-separated LSM-trees, such as TerarkDB, exhibit high performance, they experience severe space amplification. We analyze space amplification in KV-separated LSM-trees and seek efficient mitigation strategies.

\subsection{Root Causes of High Space Amplification}
While the primary purpose of GC operations is to mitigate space amplification by reclaiming space, the persistently high space amplification indicates that inefficient GC operations directly contribute to space amplification \cite{xanthakis2021parallax}. Moreover, we identify a previously overlooked aspect of space optimization: delayed compaction in the index LSM-tree.

\subsubsection{Inefficient GC operations of Value Data}
We first analyze the GC workflow and then examine the sources of overhead. We chose TerarkDB and Titan, two widely used KV-separated LSM-trees \cite{li2021differentiated, li2021elastic}, as case studies.

\textbf{Workflow of GC operations}. The GC of KV-separated LSM-trees follows the workflow described in Section \uppercase\expandafter{\romannumeral2.B}, as shown in Figure 2. Since WiscKey and Titan store value addresses within the index LSM-tree, their GC consists of four key steps: \textbf{(1) Read}: Load candidate data; \textbf{(2) GC-Lookup}: Query the index LSM-tree using candidate keys and compare the stored addresses with the scanned data to verify validity; \textbf{(3) Write}: Rewrite valid data to a new file; \textbf{(4) Write-Index}: Update the index LSM-tree to reflect the new addresses. In contrast, TerarkDB stores file numbers in the index LSM-tree and maintains file inheritance relationships, removing the necessity for index write-back. As a result, its GC comprises three steps, namely, \textbf{Read}, \textbf{GC-Lookup}, and \textbf{Write}.

\textbf{GC bottlenecks under various workloads}. To identify the root cause of inefficient GC, we conducted a latency breakdown of critical GC steps for TerarkDB and Titan. Our experiments began by loading a unique 100GB dataset, followed by updating 300GB to trigger frequent GC. We measured the total latency and execution time for each GC step, computed the proportion of each step's latency, and determined the average latency. Our tests were designed to examine latency distribution across various value lengths, covering both fixed-length and variable-length workloads. The fixed-length workloads consist of six groups with sizes ranging from 1KB to 32KB. The variable-length workloads encompassed two scenarios: one was a mixed workload with equal proportions of large (16KB) and small (100B–512B) values, and the other followed a generalized Pareto distribution \cite{hosking1987parameter, rocksdb-trace}, with average key-value sizes of 8KB and 1KB, respectively. Notably, the mixed workload is a typical pattern in ByteDance's internal OLTP database, where large-value pairs originate from writes to original data pages, while small-value pairs result from incremental write operations.

\begin{figure}[tbp]
    \setlength{\abovecaptionskip}{5pt}
    \setlength{\belowcaptionskip}{-0.6cm}
    \centering
    \includegraphics[width=0.485\textwidth]{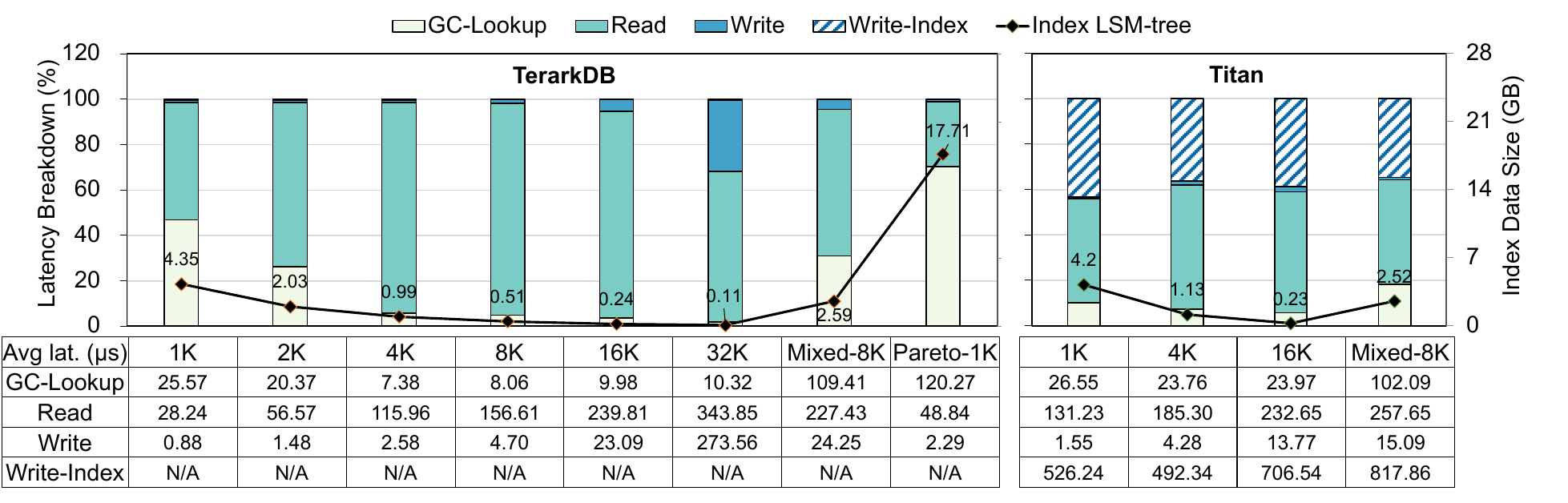}
    \caption{GC latency breakdown of TerarkDB and Titan.}
    \label{fig:mot4-gc-lat}
    \vspace{-0.5cm}
\end{figure}

GC bottlenecks under various workloads. Figure~\ref{fig:mot4-gc-lat} illustrates the relationship between GC performance bottlenecks and workloads. Specifically, \textbf{Read} incurs the highest overhead, except for the Pareto-1K. Conversely, \textbf{Write} incurs significant overhead, primarily in large-value workloads, such as the fixed-length 32K. Moreover, \textbf{GC-Lookup} incurs substantial overhead in small-value and variable-length workloads. For Titan, \textbf{Write-Index} incurs significant overhead, accounting for approximately 38\% of total GC overhead, followed by substantial \textbf{Read} overhead. If \textbf{Write-Index} were eliminated, as in TerarkDB, the overhead percentages for \textbf{Read} and \textbf{GC-Lookup} in Mixed-8K would increase by 25\% and 10\%, respectively, making them the primary performance bottlenecks. Notably, TerarkDB's average \textbf{Read} latency is significantly lower than that of Titan, as it leverages caching more effectively.

The overhead associated with GC \textbf{Read} and \textbf{Write} operations exhibits a strong positive correlation with value size, suggesting that excessively large key-value pairs substantially increase GC overhead. Additionally, we observed that \textbf{GC-Lookup} overhead also exhibits a positive correlation with the size of the index LSM-tree. Our tests revealed that the index LSM-tree is notably larger when handling small-value and variable-length workloads, resulting in increased latency, particularly for Pareto-1K workloads. Notably, although Mixed-8K and Fixed-2K exhibit similar index LSM-tree sizes and GC-Lookup latency proportions, the average GC-Lookup latency in Mixed-8K is 5.4× higher than in Fixed-2K. Further analysis reveals that this discrepancy stems from the mixed storage layout of small values and indexes in Mixed-8K, leading to a ~22\% difference in the cache hit ratio.

\subsubsection{Delayed Compactions of Index LSM-tree}
Although the substantial I/O overhead of GC is the primary driver behind excessive space amplification in KV-separated LSM-trees, the proportion of space reclaimed during each GC process also significantly affects the overall space efficiency. To address this, we began by modeling and analyzing the components of space amplification in KV-separated LSM-trees and sought to identify an additional source of space amplification that has been largely overlooked in existing research: the delayed compaction process in the index LSM-tree.

\begin{figure}[tp]
    \setlength{\abovecaptionskip}{-2pt}
    \setlength{\belowcaptionskip}{-0.6cm}
    \centering
    \includegraphics[width=0.371\textwidth]{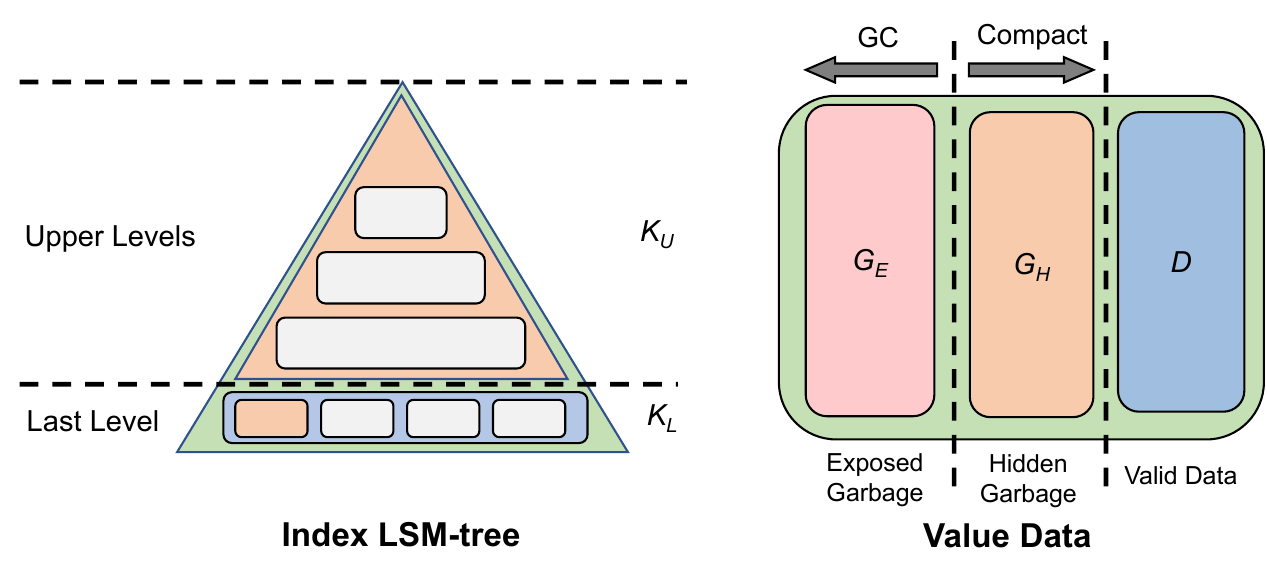}
    \caption{Space amplification analysis.}
    \label{fig:mot5-space-model}
    \vspace{-0.2cm}
\end{figure}

\begin{table}[t]
\centering
\setlength{\abovecaptionskip}{0cm}
\setlength{\belowcaptionskip}{-0.1cm}
\caption{Frequently Used Notations}
\begin{tabular}{ll}
\toprule
Notations & Descriptions \\
\midrule
$L$           & Number of levels of the index        \\
$T$           & Size factor of two neighboring levels  \\
$K_U$, $K_L$  & Size of upper and last level of the index \\
$G_E$, $G_H$  & Size of exposed and hidden garbage \\
$D$           & Size of valid data      \\
$S_{Index}$   & Space amplification of the index LSM-tree            \\
$S_{value}$   & Space amplification of the value store               \\
$P_{index}$   & Space pressure of the index LSM-tree                       \\
$P_{value}$   & Space pressure of the value store                          \\
$R_G$         & Garbage ratio threshold to trigger GC            \\
$Max_{GC}$    & Max GC threads                                   \\
$N_{threads}$ & Max threads for GC and compaction    \\
\bottomrule
\end{tabular}
\label{tab:1}
\vspace{-0.5cm}
\end{table}

\textbf{Composition of Space Amplification}. We begin by analyzing and quantifying space amplification within vanilla LSM-trees, using RocksDB as a representative case study. We activate Dynamic Capacity Adaptation (DCA) in RocksDB \cite{dynamic-level}. Once the LSM-tree stabilizes, the size of RocksDB's last level can be considered a reliable approximation of the user data set's size \cite{dayan2022spooky}. By dividing the total size of the LSM-tree by the size of the last level, we derive an approximation of space amplification \cite{dynamic-level}. As Figure \ref{fig:mot5-space-model} shows, we denote the size of the last level as $K_L$ and the total size of upper levels as $K_U$ so that we can calculate the space amplification as:
\begin{equation}
    S_{Index} \approx \frac{K_U+K_L}{K_L}=\frac{K_U}{K_L}+1
    \label{eq:lsm_sa}
\end{equation}
Empirical tests demonstrate that RocksDB exhibits a 1.11x space amplification when level multiplier \textit{T=10}. For clarity, all parameters used in this paper are summarized in Table \ref{tab:1}.

For KV-separated LSM-trees, we simplify the calculation by approximating the overall space amplification as that of the value data, since the value data size typically exceeds that of the index LSM-tree. The space amplification of value data is primarily determined by the proportion of garbage data. From the user's perspective, older versions of a key-value pair in KVS are considered garbage once the pair is updated or deleted. However, for LSM-trees using out-of-place updates, garbage data is only detectable during compactions involving multiple versions of the same key-value pair. Thus, the value data can be categorized into the following components:
\begin{itemize}
\item \textbf{Valid Data $D$} corresponds to the total size of the value data in the last level of the index LSM-tree and also represents the size of the entire unique dataset.
\item \textbf{Hidden Garbage $G_H$} refers to value data associated with old keys in the last level of the index LSM-tree awaiting compaction with the upper levels. Though invalid to users, it stays undetected until compaction. While such garbage may exist at upper levels, we focus on the last level, where most data resides and obsolete entries persist longest, making it the main source of hidden garbage.
\item \textbf{Exposed Garbage $G_E$} consists of value data associated with old keys merged in prior compactions and awaits GC operations for space reclamation.
\end{itemize}

Based on this categorization, we understand that \textbf{Hidden Garbage} arises from delayed merges of the index LSM-tree. The proportion of \textbf{Hidden Garbage} corresponds to the ratio between the upper-level data size and the size of the last level in the index LSM-tree, which can be expressed as:
\begin{equation}
\frac{G_H}{D} \approx
\frac{K_U}{K_L}
\label{eq:ratio}
\end{equation}

Based on the analysis above, the space amplification in KV-separated LSM-trees ($S_{value}$) can be expressed as
\begin{equation*}
S_{value} = \frac{G_E}{D}+(\frac{G_H}{D}+1) \approx \frac{G_E}{D}+(\frac{K_U}{K_L}+1)
\label{eq:sa}
\end{equation*}
\begin{equation}
S_{value} \approx \frac{Exposed \ Garbage}{Valid \ Data}+S_{Index}
\label{eq:sa2}
\end{equation}

Specifically, the size of \textbf{Valid Data $D$} depends on the user data set size. Both \textbf{Hidden Garbage $G_H$} and \textbf{Exposed Garbage $G_E$} contribute collectively to space amplification. The size of \textbf{Exposed Garbage} is mainly influenced by GC operations. On the other hand, the size of \textbf{Hidden Garbage} is primarily determined by the space amplification of the index LSM-tree, which is influenced by compaction operations.

\begin{figure}[tbp]
    \setlength{\abovecaptionskip}{5pt}
    \setlength{\belowcaptionskip}{-0.6cm}
    \centering
    \subfloat[Space amplification of index]{
            \includegraphics[width=.47\columnwidth]{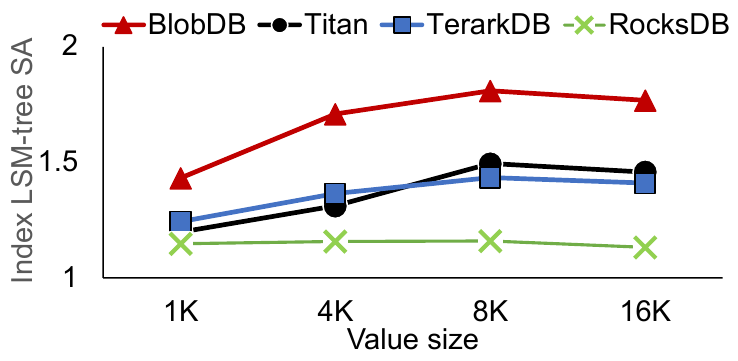}
            \label{fig:mot2-1}
        }
    \subfloat[Exposed garbage of value]{
            \includegraphics[width=.47\columnwidth]{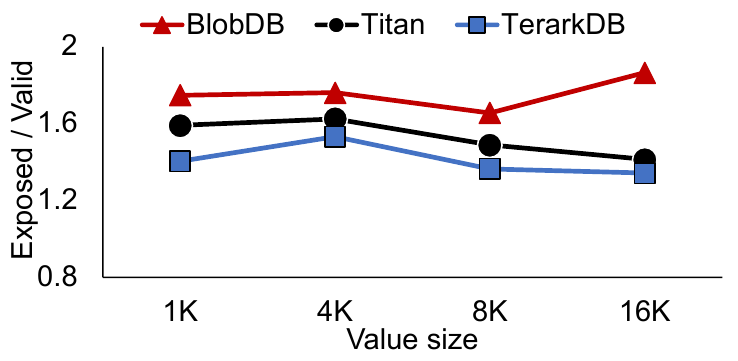}
            \label{fig:mot2-2}
        }
    \caption{Two sources of space amplification.}
    \label{mot2}
    \vspace{-0.5cm}
\end{figure}
To quantify the impact of the index LSM-tree on space amplification, we measured both its space amplification and the \textbf{Exposed/Valid} ratio in the value store. As shown in Figure~\ref{mot2}, the index LSM-tree exhibits significantly higher space amplification than RocksDB. Under a Fixed-8K workload in TerarkDB, delayed compaction in the index LSM-tree prevented the reclamation of about 51.2\% of the garbage (i.e., \textbf{Hidden Garbage}) in the value store, while the remaining 48.8\% (i.e., \textbf{Exposed Garbage}) was reclaimable. Similar patterns were observed across other KV-separated LSM-trees.

\textbf{Delayed Compactions of Index LSM-tree}. To further investigate the reasons behind the space amplification of the LSM-tree index compared to RocksDB, we analyzed both the tree structure of the index LSM-tree and the frequency of compaction operations. For instance, under the Fixed-8K workload, we observed that the KV-separated TerarkDB flushed numerous small-sized kSSTs to disk. These kSSTs had an average size of only 211K, much smaller than the default 64M SST size in vanilla LSM-trees. Smaller SST sizes hinder reaching the level size trigger threshold, delaying compaction execution. We collected statistics on the number of compaction executions. Under the Fixed-8K workload with KV separation, we found that the number of compaction executions was only 27.8\% of those in a vanilla LSM-tree. Additionally, the smaller SSTs result in fewer levels in index LSM-trees, typically about 2-3 levels, compared to 5 levels in vanilla LSM-trees under Fixed-8K. Fewer levels complicate maintaining fixed-level ratios and reduce compaction parallelism. As a result, a significant amount of data accumulates in the intermediate levels without efficient downward progression through compaction, leading to worsening space amplification.

\section{Design}
In this section, we introduce \textbf{Scavenger+}, a new KV-separated LSM-tree-based key-value store. Compared to other KV-separated LSM-trees, Scavenger+ enhances foreground performance under equivalent space constraints, achieves lower space amplification than existing solutions, and thus provides a better trade-off between performance and space.

\subsection{System Overview}
\begin{figure}[tbp]
    \setlength{\abovecaptionskip}{5pt}
    \setlength{\belowcaptionskip}{-0.6cm}
    \centering
    \includegraphics[width=0.42\textwidth]{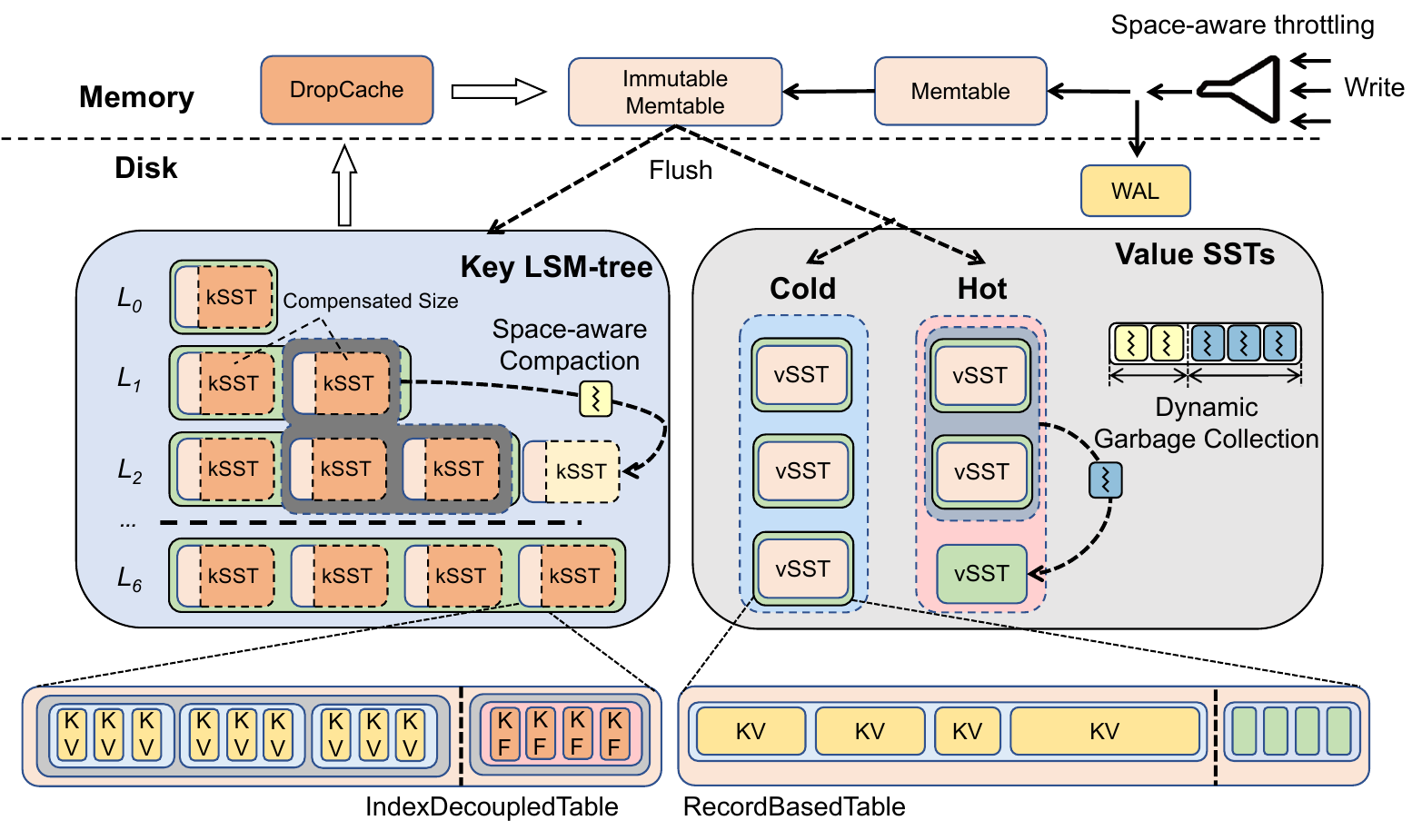}
    \caption{Overview of Scavenger+.}
    \label{fig:design1-all}
    \vspace{-0.5cm}
\end{figure}
Scavenger+ employs Memtables to buffer user writes and a Write-Ahead Log (WAL) to ensure consistency, similar to most LSM-trees. It performs KV separation during flushing to enhance range query performance and support transactions, akin to industrial KV-separated LSM-trees. The primary distinction from other schemes lies in Scavenger+'s refinement of the garbage collection (GC) and compaction processes to provide an optimal trade-off between execution time and space, necessitating several new components, as shown in Figure~\ref{fig:design1-all}. Specifically, Scavenger+ introduces \textbf{RecordBasedTable} and \textbf{IndexDecoupledTable} to store values and indexes, replacing the original \textbf{BlockBasedTable} to reduce I/O operations during GC-Read and GC-Lookup. Concurrently, Scavenger+ introduces \textbf{DropCache} to identify hot data and employs a lightweight strategy that is aware of data hotspots for hot-cold data separation. Furthermore, Scavenger+ implements an adaptive prefetching strategy for GC operations to further reduce the number of I/O operations and accelerate GC execution. Scavenger+ uses a space-aware compaction strategy based on compensated size to minimize space amplification in the index LSM-tree. Finally, Scavenger+ introduces dynamic GC scheduling for dynamic resource allocation between compaction and GC, thus quickly reclaiming space.

\subsection{I/O Efficient Garbage Collection}
Our findings from the evaluation in Section \uppercase\expandafter{\romannumeral2}.D indicate that critical steps in GC (\textbf{Read}, \textbf{GC-Lookup}, and \textbf{Write}) act as bottlenecks affecting GC efficiency. To address these bottlenecks, we propose an I/O-efficient GC scheme. This scheme aims to reduce the I/O overhead of each step, thereby speeding up GC execution and enhancing space reclamation.

\subsubsection{Lazy Read of the GC} 
As depicted in Figure~\ref{fig:mot4-gc-lat}, GC-Read is the primary contributor to latency for workloads dealing with large values. The latency of GC-Read increases with larger average value sizes due to increased I/O accesses. GC-Read does not need to access the value data of every key-value pair. During GC execution, only valid key-value pairs are copied and rewritten, while invalid ones are discarded to reclaim storage space. Since GC cannot predict valid key-value pairs beforehand, it requires a validity check (GC-Lookup) for each pair. GC-Lookup essentially performs point queries on specific keys, relying on keys read during GC-Read. Unordered value logs and block-based tables do not support both lazy value reading and fast key retrieval. To address this, we designed a new table structure for value SSTables.

\begin{figure}[tbp]
    \setlength{\abovecaptionskip}{5pt}
    \setlength{\belowcaptionskip}{-0.6cm}
    \centering
    \subfloat[Storage layout of RTable.]{
            \includegraphics[width=.47\columnwidth]{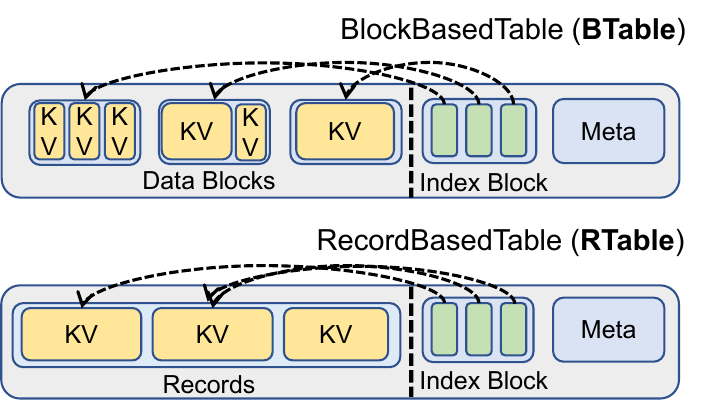}
            \label{fig:design2-read}
        }
    \subfloat[Lazy Read process of GC.]{
            \includegraphics[width=.46\columnwidth]{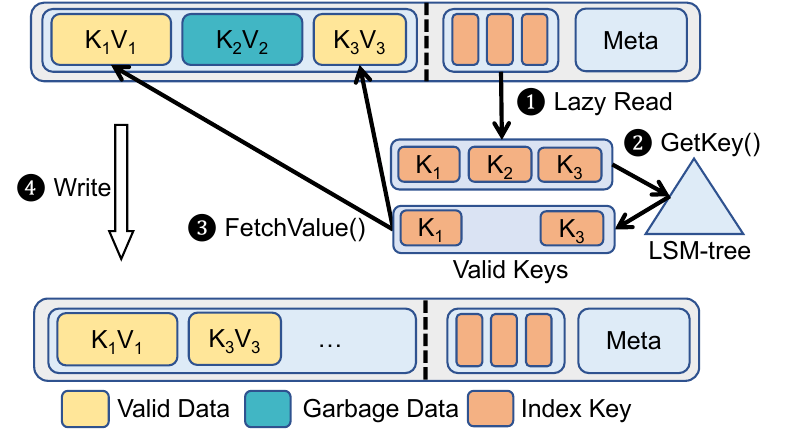}
            \label{fig:design3-read}
        }
    \caption{Lazy Read with RTable.}
    \label{des2}
    \vspace{-0.5cm}
\end{figure}

\textbf{Storage layout of vSST}. As illustrated in Figure 4, the unordered value log of WiscKey stores key-value pairs in an unordered manner. This structure hinders quick access to keys and leads to suboptimal query performance. The BlockBasedTable (\textbf{BTable}) used by RocksDB and TerarkDB provides dedicated index blocks but maintains a sparse index for data blocks, preventing access to all keys solely through these index blocks. Moreover, the block-based layout hinders the lazy reading of individual key-value pairs. To address these limitations, Scavenger+ introduces a new table structure for value SSTables, RecordBasedTable (\textbf{RTable}). Specifically, RTable organizes key-value pairs into records sequentially and constructs a \(\langle \text{key}, \text{offset} \rangle\) tuple for each record, storing it in the index block as an index entry. Other metadata structures, like Footer, Meta Index, and Filter, resemble those in BTable. Fundamentally, RTable maintains a dense index for key-value pairs, allowing GC-Read to access all keys directly through the index block, removing the need to access values every time. After verifying a key-value pair, GC-Read can lazily read the value since the corresponding index entry directly pinpoints it. With this design, foreground reads can directly locate key-value pairs, avoiding the retrieval of data blocks.

Notably, RTable increases the index size compared to BTable by maintaining a dense index for each key-value pair. Under Pareto-1K, RTable introduces approximately 2\% additional storage overhead, which is negligible compared to the dataset. Setting a higher threshold for key-value separation or constructing RTable only for vSSTs with an average size greater than 1K can help reduce additional space. To mitigate I/O overhead, we employ a partitioned index \cite{part-index}, which allows read requests to load only the necessary index blocks.

\textbf{Lazy Read}. RTable enables GC threads to perform GC-Lookup directly after reading keys, delaying the reading of values until GC-Lookup confirms the validity of the key-value pairs. This process, known as Lazy Read, minimizes I/O overhead during GC-Read. Figure~\ref{fig:design3-read} shows a GC example with Lazy Read enabled. Initially, the GC thread reads the index block of the selected vSST, retrieves keys $K_1$ to $K_4$, and stores them in the high-priority cache. Then, it calls the $GetKey()$ function to perform GC-Lookup, checking the corresponding keys' validity. Key-value pairs of $K_1$ and $K_3$ are identified as valid and need to be copied and rewritten. Consequently, the GC thread uses existing address information to read $K_1V_1$ and $K_3V_3$ values, appending them to the new vSST, thus eliminating access to invalid key-value pair values.

\subsubsection{Index-Record Separation}
Lazy Read mitigates GC-Read overhead in workloads with large values. However, Figure~\ref{fig:mot4-gc-lat} shows that GC-Lookup is a major contributor to GC overhead, especially in workloads with small fixed-length and variable-length values. Since GC-Lookup essentially performs point queries on the index LSM-tree, optimizations such as caching and indexing can enhance its performance. We aim to explore how KV-separated LSM-trees can better leverage existing optimizations and address unique challenges.

The experimental results in Figure~\ref{fig:mot4-gc-lat} reveal that the LSM-tree index sizes for fixed-length (Fixed-1K, Fixed-2K) and variable-length workloads (Mixed-8K, Pareto-1K) exceed the pre-allocated cache size of 1GB (1\% of the dataset). While allocating extra memory for caching may be beneficial, practical resource quotas often constrain usable memory, particularly in high-density server environments. GC-Lookup, though fundamentally a point query, does not require access to all data within the index LSM-tree, particularly excluding KV pairs with values below the KV separation threshold. Specifically, GC-Lookup only examines indexes for KV pairs exceeding the KV separation threshold to validate their existence. However, current implementations often store small KV pairs and indexes together in the same data block, forcing GC-Lookup to access small KV pairs, leading to unnecessary I/O. To address this challenge, Scavenger+ introduces a novel table structure that enhances index and small KV pair separation.

\begin{figure}[tbp]
    \setlength{\abovecaptionskip}{5pt}
    \setlength{\belowcaptionskip}{-0.6cm}
    \centering
    \subfloat[Storage layout of DTable.]{
            \includegraphics[width=.41\columnwidth]{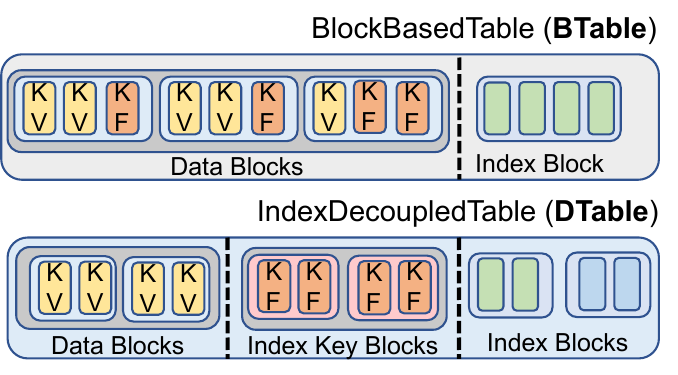}
            \label{des5-dtable}
        }
    \subfloat[GC-Lookup process.]{
            \includegraphics[width=.52\columnwidth]{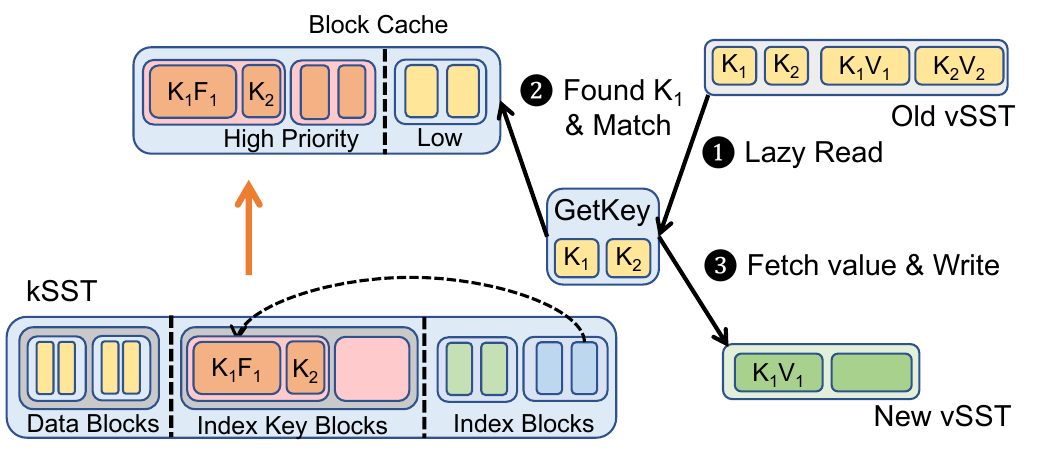}
            \label{fig:design5-discuss}
        }
    \caption{GC-Lookup with DTable}
    \label{des3}
    \vspace{-0.5cm}
\end{figure}

\textbf{Storage layout of kSST.}. Scavenger+ aims to facilitate rapid access to standalone indexes and reduce unnecessary I/O overhead. To this end, we propose a cache-efficient table structure, \textbf{IndexDecoupledTable} (\textbf{DTable}). Specifically, as depicted in Figure \ref{des5-dtable}, unlike the traditional BTable, which interleaves small key-value (\textbf{KV}) records and indexes (\(\langle \text{key}, \text{file\_number} \rangle\), \textbf{KF}), DTable separates small KV records and index items into distinct data and index blocks, storing them separately. This design allows small records (KVs) and index items (KFs) to be accessed independently, allowing GC-Lookup to directly access index key blocks without extra I/Os. Additionally, Scavenger+ assigns a higher cache priority to index key blocks as they contain more entries than regular data blocks and support multiple operations, including foreground queries and GC-Lookup. A higher cache priority ensures that index key blocks are retained in the block cache's high-priority queue \cite{rocksdb-bcache}, where data remains longer before eviction.

Notably, DTable’s design remains fully compatible with BTable. It retains BTable’s structure when all key-value pairs fall into the same size category (either large or small). However, for variable-length workloads, it reduces GC-Lookup I/O overhead by storing indexes separately. In foreground queries, DTable rarely incurs additional I/O due to overlapping ranges. This occurs because Scavenger+ assigns a higher cache priority to index key blocks and implements proactive cache replacement during compaction. This approach can eliminate additional index I/O for foreground queries on large key-value pairs, further improving read performance.

\textbf{GC-Lookup} often exhibits strong access locality as it performs multiple point queries within a continuous key range. As shown in Figure \ref{fig:design5-discuss}, after the GC thread acquires keys $K_1$ and $K_2$, it initiates GC-Lookup. If a cache miss occurs while querying $K_1$, the corresponding index key blocks are loaded and inserted into the high-priority queue of the block cache, followed by internal block verification. If valid, the corresponding $K_1V_1$ is then read and written to the new vSST. While verifying $K_2$, GC-Lookup directly hits the block cache, thus accelerating GC-Lookup and reducing unnecessary I/O.

\subsubsection{Hotspot-aware writing}
As GC-Read and GC-Lookup overheads decrease, GC-Write overhead becomes more dominant, especially under large-value workloads. Minimizing the amount of valid data rewritten during GC can improve efficiency. At its core, GC reclaims space occupied by garbage generated by repeated user writes or delete operations, which primarily consist of hot write data, where deletions are treated as overwrites to existing data \cite{zhang2022halsm}. A higher proportion of hot write data in a single file typically correlates with higher garbage accumulation, allowing GC to reclaim more storage space. Therefore, to fully exploit the inherent hot and cold data characteristics of workloads \cite{cao2020characterizing}, Scavenger+ proposes a hotspot-aware writing strategy to identify data hotspots, leveraging this insight to optimize Flush and GC operations.

\textbf{Hotspot Identification}. Prior approaches, such as TRIAD \cite{balmau2017triad}, identify hot keys by selecting top-K entries with the highest update frequencies in memory. During flush, hot entries are retained in memory while cold entries are flushed to disk. Scavenger+ similarly leverages write locality but infers hot keys by tracking overwritten/deleted keys during compaction, storing recently detected hot write keys in an LRU-based DropCache. As hot data typically occupies a small portion of the dataset, DropCache ensures low memory overhead by storing only recently identified hot write keys (32B per KV). For larger datasets or higher hot-data ratios, Scavenger+ can further optimize memory usage by employing probabilistic data structures, such as Cuckoo Filters \cite{fan2014cuckoo}. This design enables hot-cold separation in a lightweight manner, making it particularly suitable for write-intensive workloads where compaction events naturally reveal access patterns.

\textbf{Hotspot-aware Flush and GC}. Leveraging hotspot identification, Scavenger+ implements a hotspot-aware write strategy for vSST generation, specifically optimizing Flush and GC operations. As illustrated in Figure~\ref{fig:design1-all}, Scavenger+ determines if each key-value pair is present in DropCache during vSST writing. If a key is found in DropCache, it is classified as hot data and written to the hot vSST; otherwise, it is written to the cold vSST. After completing Flush and GC, Scavenger+ segregates data into hot and cold vSSTs. Hot vSSTs contain a larger proportion of hot write data, exhibiting a higher garbage ratio than cold vSSTs. Our GC strategy employs a garbage ratio-based greedy algorithm, prioritizing high-garbage-ratio files for GC. This strategy increases the probability of GC on hot vSSTs, thus minimizing unnecessary GCs on cold data. 

\begin{figure}[tbp]
    \setlength{\abovecaptionskip}{5pt}
    \setlength{\belowcaptionskip}{-0.6cm}
    \centering
    \includegraphics[width=0.47\textwidth]{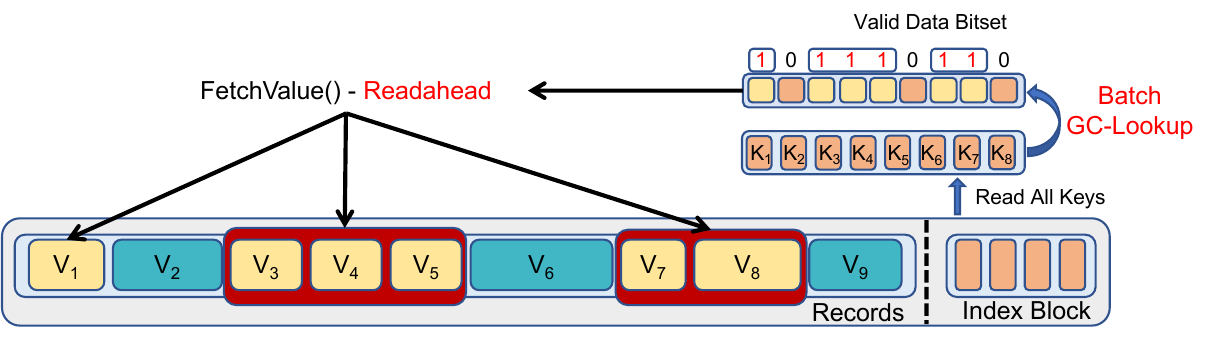}
    \caption{Adaptive Readahead for GC}
    \label{fig:design-ext-readahead}
    \vspace{-0.5cm}
\end{figure}

\subsubsection{Adaptive Readahead for GC}
RTable enables GC-Read to access all keys and valid values, thereby reducing read overhead from invalid pairs. The GC lazy read process for valid key-value pairs involves two steps: \textbf{ReadKey} and \textbf{ReadValue}. However, GC-Read still triggers multiple I/O operations because ReadKey, GC-Lookup, and ReadValue are executed serially for each key-value pair. With a low garbage ratio, indicating a high proportion of valid data in vSST, execution can be accelerated using the readahead technique, which features larger I/O sizes and fewer operations. However, the mixed storage layout of valid and invalid data complicates determining the optimal readahead I/O size. To address this, Scavenger+ introduces adaptive readahead for GC lazy read, further reducing I/O overhead.

After reading all keys via the RTable, GC-Lookup verifies the validity of all key-value pairs in the vSST. Scavenger+ replaces serial execution with a batch GC-Lookup approach. During execution, Scavenger+ generates a valid data bitmap to record verification results, indicating each key-value pair's validity in the vSST. It identifies contiguous segments of continuously stored valid key-value pairs and calculates the readahead I/O size by multiplying the segment size by the average key-value pair size in the vSST. It then performs readahead on each valid data segment individually, effectively reducing I/Os without incurring I/O waste.

Figure~\ref{fig:design-ext-readahead} illustrates the adaptive readahead process in Scavenger+ with an example. Upon retrieving keys $K_1$ to $K_8$, the batch lookup yields a valid data bitmap, revealing that $K_3$ to $K_5$ and $K_6$ to $K_7$ form two consecutive segments with readahead sizes of 3 and 2, respectively. Assuming an average key-value pair size of 4KB in the vSST, the readahead I/O sizes are 12KB and 8KB, respectively. Compared to GC-Read without adaptive readahead, this method reduces I/Os from 6 to 3, thus enhancing I/O efficiency.

\subsection{Space-aware Compaction based on Compensated Size}
\begin{figure}[tbp]
        \setlength{\abovecaptionskip}{5pt}
        \setlength{\belowcaptionskip}{-0.6cm}
        \centering
        \subfloat[Vanilla file size]{
            \includegraphics[width=.33\columnwidth]{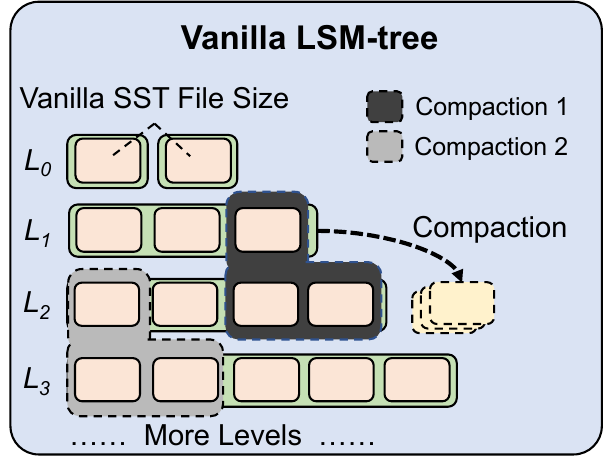}
            \label{fig:des2-comp-va-file-size}
        }\hspace{-3mm}
        \subfloat[kSST file size]{
            \includegraphics[width=.273\columnwidth]{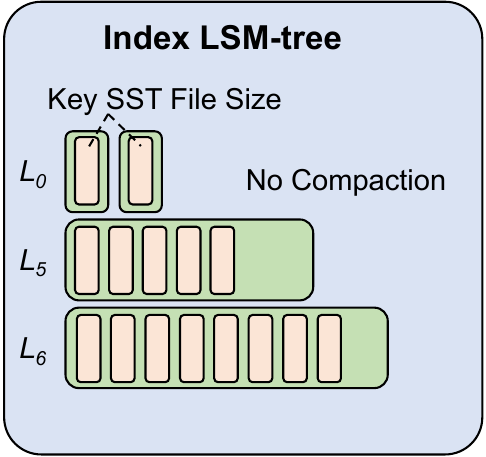}
            \label{fig:des2-comp-file-size}
        }\hspace{-3mm}
        \subfloat[Compensated file size]{
            \includegraphics[width=.33\columnwidth]{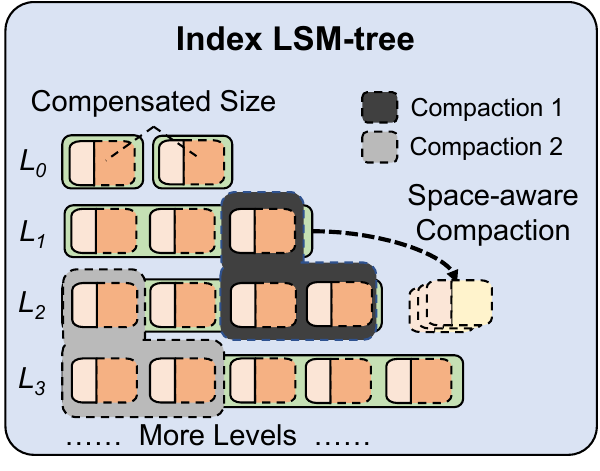}
            \label{fig:des2-comp-csize}
        }
        \caption{Different compaction strategies.}
        \vspace{-0.5cm}
\end{figure}

The index LSM-tree contributes to the space amplification in KV-separated LSM-trees (detailed in Section \uppercase\expandafter{\romannumeral2}.D). This is primarily because, after KV separation, the reduced SST file sizes delay the index LSM-tree's compaction operations. As depicted in Figure~\subref*{fig:des2-comp-file-size}, the smaller SST sizes make it harder to reach the compaction trigger threshold. Furthermore, the reduced data volume in the index LSM-tree hinders its ability to form a multi-level structure like traditional LSM-trees, as illustrated in Figure~\subref*{fig:des2-comp-va-file-size}. This hindrance to parallel compaction reduces the likelihood of triggering intermediate-level compaction, significantly contributing to space amplification. Prompt compaction scheduling is crucial; without it, substantial data accumulates in the upper levels, preventing space amplification from converging to 1.11x. Therefore, for optimal space amplification, KV-separated LSM-trees must promptly schedule compaction to maintain a standard multi-level structure with a fixed inter-level ratio of 10.

A straightforward approach involves lowering compaction trigger thresholds to enable more frequent compaction. However, determining suitable thresholds and adapting to workload fluctuations pose significant challenges. Vanilla LSM-trees, such as RocksDB, effectively control space amplification through leveled compaction and dynamic leveling strategies \cite{dynamic-level}. Essentially, these mechanisms are designed to account for actual space usage at each level. However, after KV separation, the index LSM-tree, which stores only indexes, is significantly reduced in size and can no longer reflect actual space usage. At this point, space amplification in the index LSM-tree directly affects that of the value data. Scavenger+ addresses this issue by introducing a space-aware compaction strategy.

\textbf{Compaction strategy based on compensated size} considers the original data size before scheduling compaction. Specifically, it calculates the compaction score for each level by considering the actual size of the value associated with the index, treating this as the logical size for that level. This strategy effectively transforms a separated LSM-tree into a non-separated one, thus ensuring prompt compaction scheduling. The compensated size of the kSST can be obtained either by storing the actual size of the associated key-value pairs in the index LSM-tree, as done in BlobDB and Titan, or by calculating the average size of the associated vSST for each kSST based on the dependencies between kSSTs and vSSTs, as in TerarkDB. Furthermore, as the compensated size calculation relies on the actual value size, it is not dependent on a specific type of workload. It adapts well to workloads with varying value sizes, outperforming static configurations. Additionally, a larger compensated size for a kSST indicates that its corresponding actual data size occupies more storage space. Consequently, the kSST file with the maximum compensated size is selected to initiate compaction.

Compensation enables the index LSM-tree to quickly form a multi-level structure akin to that of the vanilla LSM-tree by facilitating frequent compaction, thereby constructing more levels. Increased levels often result in reduced data dependency between compaction jobs, enabling parallel compaction execution and further accelerating compaction, as depicted in Figure~\subref*{fig:des2-comp-csize}. The standard multi-level structure and the fixed inter-level ratio (e.g., 10) guarantee that the index LSM-tree's space amplification swiftly converges to the ideal value of 1.11x, akin to RocksDB, as illustrated in Figure~\subref*{fig:test6-features-index-sa}, thereby reducing the overall space amplification. However, it is noteworthy that reducing the index LSM-tree's space amplification primarily improves the whole system's garbage ratio, and effectively reducing space amplification necessitates completing GC operations for space reclamation.

\subsection{Dynamic GC Scheduling}
Upon identifying the two sources of space amplification, Scavenger+ proposes I/O-efficient GC and space-aware compaction to optimize each source individually. However, in resource-limited scenarios, competition often arises for CPU and storage resources between the two key operations: GC and compaction. These operations can also compete with foreground user threads, thus affecting foreground performance. Furthermore, while GC operations primarily reduce the system's overall space amplification, the space-aware compaction strategy enhances GC efficiency, enabling a single GC operation to reclaim more storage space. Consequently, more judicious scheduling of GC and compaction is necessary to achieve improved space-time trade-offs. To this end, we propose a dynamic GC scheduling strategy.

\begin{figure}[tbp]
    \setlength{\abovecaptionskip}{5pt}
    \setlength{\belowcaptionskip}{-0.6cm}
    \centering
    \includegraphics[width=0.4\textwidth]{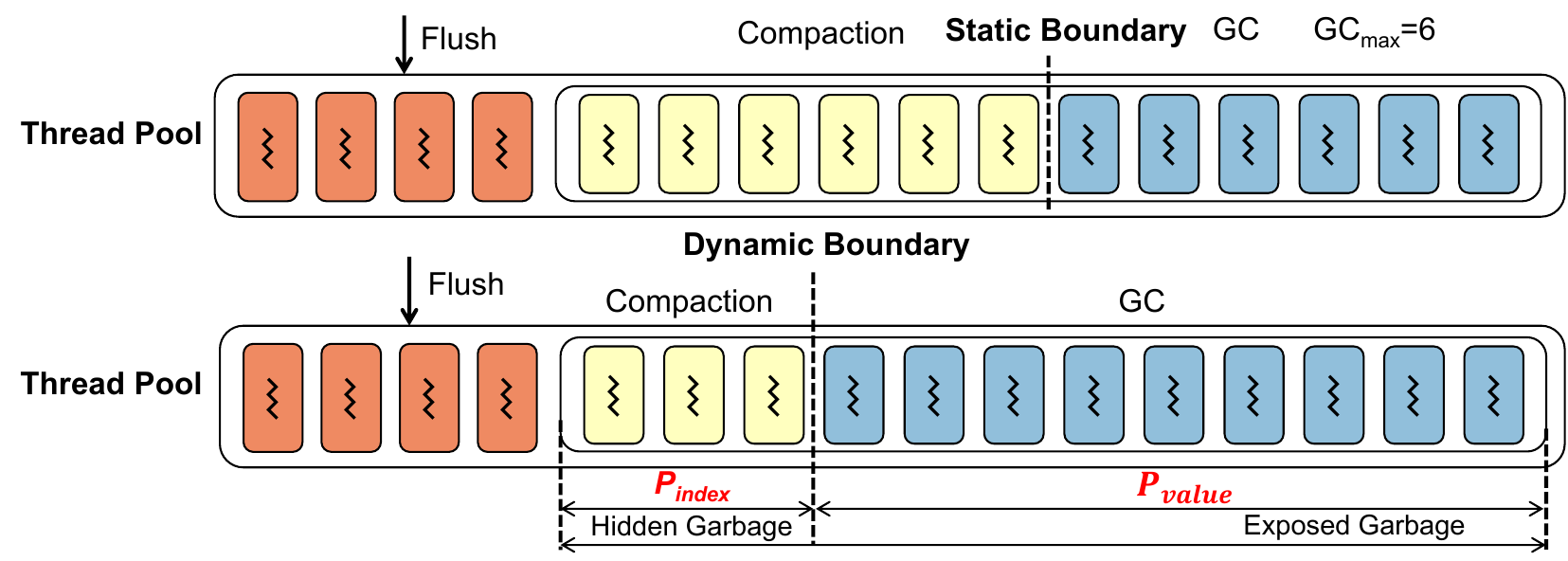}
    \caption{Dynamic thread allocation for GC}
    \label{fig:design-dyna-gc}
    \vspace{-0.5cm}
\end{figure}

\subsubsection{Dynamic thread allocation}
Compaction and GC tasks for KV-separated LSM-trees are scheduled and executed by the background task thread pool. Typically, the thread pool size is user-specified with static thread allocation. With a thread pool size of 12 for Compaction and GC, a maximum of 6 threads are allocated to GC by default, meaning only up to half of the thread resources can be used for GC \cite{terarkdb}. However, GC often experiences increased execution pressure. It becomes crucial to perform GC to reclaim storage space when space is depleted. In such cases, static thread allocation often fails to fully utilize CPU resources. Therefore, Scavenger+ introduces a dynamic thread allocation strategy considering the space amplification pressure of different components.

As shown in Figure~\ref{fig:design-dyna-gc}, Scavenger+ allocates threads based on the space amplification pressure faced by the index LSM-tree ($P_{index}$) and the value store ($P_{value}$). Specifically, the difference between the actual and ideal space amplification for the index LSM-tree, representing the change caused by compaction, serves as its space amplification pressure. The ideal space amplification of an LSM-tree, determined by the size ratio $T$ and the number of levels $L$, allows the space amplification pressure to be expressed as:
\begin{equation}
P_{index}=S_{index}-(1+\sum_{i=1}^{L-1}\frac{1}{T^i})
\end{equation}
Similarly, the ideal space amplification of value data, determined by the garbage ratio threshold $R_G$ to trigger GCs, allows the space amplification pressure to be expressed as:
\begin{equation}
P_{value}=\frac{G_{E}}{D}-\frac{R_G}{1-R_G}
\end{equation}
Based on the space pressures of the index LSM-tree and the value store, Scavenger+ employs a greedy algorithm on the available background threads ($N_{threads}$) to allocate the current maximum thread count for GC ($Max_{GC}$), expressed as:
\begin{equation}
Max_{GC}=N_{threads} * \frac{P_{value}}{P_{index}+P_{value}}
\end{equation}
With dynamic allocation, additional threads can be allocated for GC to rapidly reclaim space when the index LSM-tree's compaction pressure is low. Conversely, when compaction pressure is high, additional threads are dedicated to compaction, thereby enhancing the efficiency of subsequent GC.

Unlike scheduling approaches like SILK \cite{balmau2019silk}, which determine background tasks based on workload pressure, Scavenger+ operates without workload assumptions. Instead, it accounts for compaction and GC pressure based on the observed space amplification in the index LSM-tree and value data. This approach minimizes space costs by efficiently utilizing user-allocated thread resources. However, keeping all background threads busy may degrade foreground performance.

\subsubsection{Background bandwidth limit}
Although increasing GC concurrency indiscriminately may expedite space reclamation, GC's I/O overhead causes numerous GC threads to compete with user threads for storage bandwidth. As a result, high GC concurrency can degrade foreground read and write performance. To swiftly reclaim storage space without degrading foreground performance, Scavenger+ dynamically limits GC bandwidth to mitigate the impact on user requests.

For writes, when the LSM-tree write disk bandwidth reaches 80\% of the total bandwidth due to concurrent flush and GC tasks, Scavenger+ compares the current flush bandwidth to its average. A decrease exceeding 20\% indicates contention between GC and user writes. In such scenarios, Scavenger+ employs a RateLimiter to dynamically throttle the background GC write bandwidth, gradually decreasing it by a fixed step (e.g., 20\%) to minimize contention between GC and the user writes. A similar approach applies to reads. Scavenger+ limits the GC read bandwidth to reduce interference with user reads. By regulating the I/O bandwidth of GC, Scavenger+ minimizes the background impact on foreground performance, optimally schedules compaction and GC tasks for rapid space reclamation, and achieves better space-time trade-offs.

Notably, compaction and GC scheduling based on space amplification pressure prioritizes space amplification reduction, which is more critical than tail latency in KV separation. Additionally, Scavenger+ dynamically throttles the GC bandwidth to delay GC execution, reducing its impact on foreground performance. Unlike delayed compaction, this approach does not introduce latency spikes. Instead, it improves the efficiency of each GC operation with more garbage accumulation. By coordinating both CPU and I/O resources, Scavenger+ achieves better space-time trade-offs.

\section{Evaluation}
In this section, we evaluate Scavenger+ and compare it with several state-of-the-art KV-separated LSM-trees: BlobDB, Titan, and TerarkDB. We also tested the most popular vanilla LSM-tree, RocksDB (v8.2.1), as our baseline.

\subsection{Setup}
\textbf{Testbed}. Our test environment consists of a 32-core Intel(R) Xeon(R) Silver 4314 CPU @ 2.40GHz and 64GB DDR4 memory. In addition, we employed a 500G KIOXIA NVMe SSD. The operating system kernel for the server is Linux 5.4, with an operating system version of Debian 10. We formatted the NVMe SSD using the ext4 file system.

\textbf{System Configuration}. Using the tuning manual \cite{rocksdb-tuning}, we standardized the configurations in all the KVSs evaluated. We configured the key-value separation threshold at 512 bytes, the Memtable size at 64MB, the Key Sorted String Table (kSST) size at 64MB, the Value Sorted String Tables (vSST) size at 256MB, the BlockCache size at 1GB (approximately 1\% of the 100GB dataset), and the bloom filters at 10 bits per key. Both the foreground and background threads were set to 16. Our system employs direct I/O for background flush and compaction tasks to reduce the impact on the page cache. The garbage ratio threshold for triggering GC was established at 0.2 to control the frequency of GC operations.

\textbf{Fair Comparison}. Evaluating the foreground performance of various KV-separated LSM-trees without constraining space usage leads to unfair comparisons. We implement space-aware throttling and establish a space threshold that limits the maximum space for each KVS, ensuring a fair performance comparison and controlled space cost. Unless otherwise specified, the space limit is set at 1.5 times the size of the dataset.

\textbf{Workload}. We evaluate both fixed and variable-length workloads using a modified version of RocksDB's dbbench and YCSB-C \cite{ycsbc}, a C++ implementation of YCSB \cite{cooper2010benchmarking}. The key size is fixed at 24B. For value sizes, we evaluate both fixed-length and variable-length workloads. The variable-length workloads include: (1) Mixed-8K, which simulates the ByteDance OLTP database pattern with a 1:1 ratio of small values (uniformly distributed from 100 to 512 bytes) to large values (16KB), averaging about 8KB; and (2) Pareto-1K, where value sizes follow a generalized Pareto distribution \cite{hosking1987parameter, rocksdb-trace}, averaging about 1KB \cite{li2021differentiated}. Key distribution follows a Zipfian distribution \cite{cooper2010benchmarking}, reflecting real-world scenarios.

\subsection{Microbenchmarks}
We assessed the throughput of various operations across different workloads (Mixed-8K and Pareto-1K), including the insertion of 100GB of KV pairs and the update of 300GB, reading 300GB via point queries, and executing 40 million range query requests. For range query (scan) workloads, scan lengths ranged uniformly from 2 to 1000. The testing procedure involved randomly loading 100GB of data and executing the update workload to generate significant garbage, triggering GC operations. Subsequently, we performed the read and scan workloads. To investigate the trade-offs between space usage and time efficiency in existing schemes, we evaluated performance under a maximum space usage of 1.5 times the dataset size (150GB storage quota) and examined performance and space amplification in scenarios without space limits.

\begin{figure*}[tbp]
        \setlength{\abovecaptionskip}{5pt}
        \setlength{\belowcaptionskip}{-0.7cm}
	\centering
	\subfloat[Throughput under Mixed-8K]{
            \includegraphics[width=.30\textwidth]{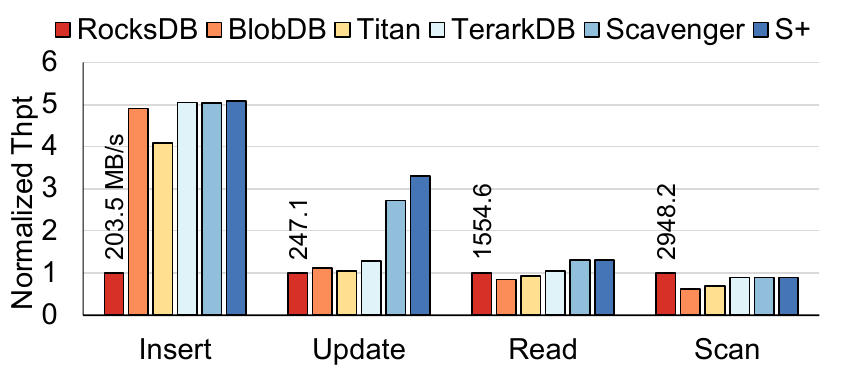}
            \label{fig:test1-micro-mixed-iops}
        }
        \subfloat[Throughput under Pareto-1K]{
            \includegraphics[width=.30\textwidth]{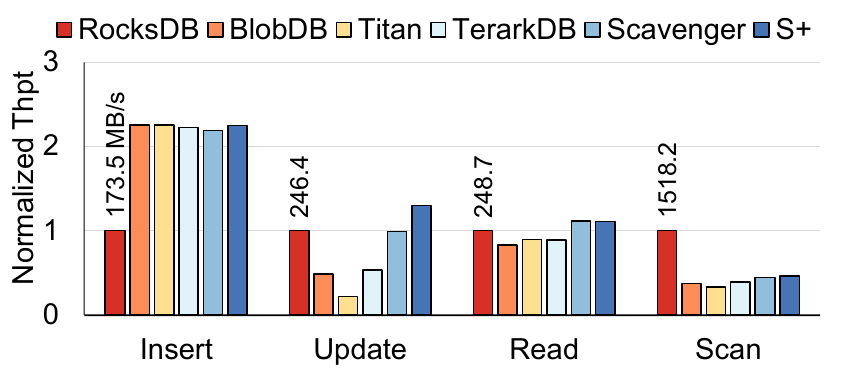}
            \label{fig:test1-micro-pareto-iops}
        }
        \subfloat[Disk I/O under Mixed-8K Update]{
            \includegraphics[width=.30\textwidth]{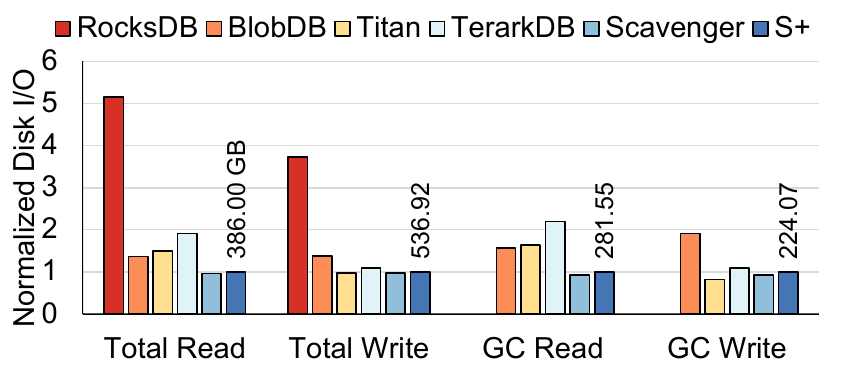}
            \label{fig:test1-micro-ndb-io}
        }
        \caption{Microbenchmarks under Mixed-8K and Pareto-1K with 1.5x space limit (\textit{\textbf{S+} denotes Scavenger+}). }
        \vspace{-0.5cm}
\end{figure*}

\begin{figure}[tbp]
        \vspace{-0.3cm}
        \setlength{\abovecaptionskip}{5pt}
        \setlength{\belowcaptionskip}{-0.6cm}
	\centering
	\subfloat[Write Throughput]{
            \includegraphics[width=.44\columnwidth]{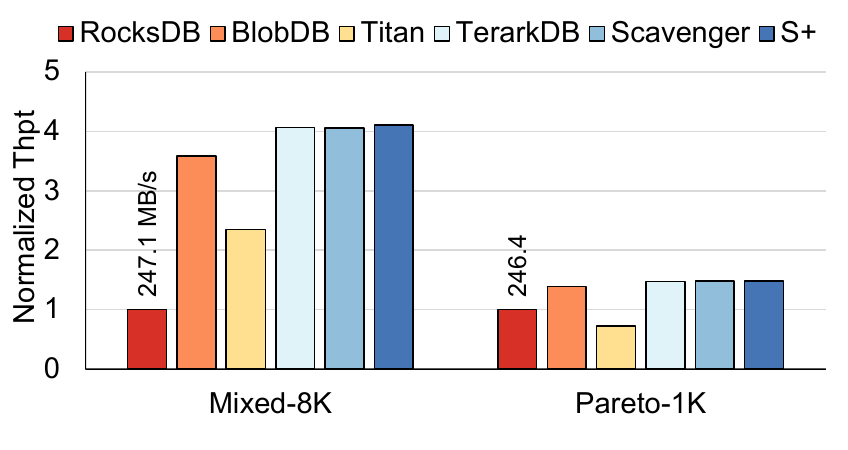}
            \label{fig:test2-micro-nolimit-update}
        }
        \subfloat[Space amplification]{
            \includegraphics[width=.44\columnwidth]{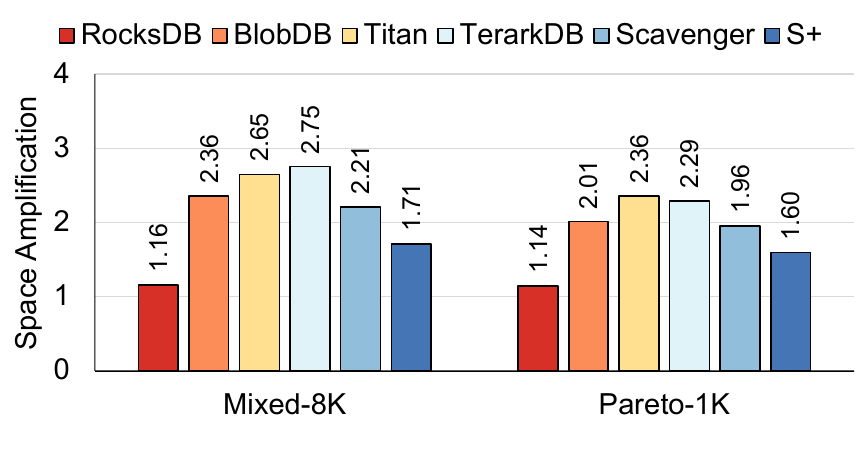}
            \label{fig:test2-micro-nolimit-space}
        }
	\caption{Microbenchmarks without space limits.}
        \label{fig:test2-micro-nolimit}
        \vspace{-0.2cm}
\end{figure}

\begin{figure}[tbp]
        \setlength{\abovecaptionskip}{0pt}
        \setlength{\belowcaptionskip}{-0.6cm}
	\centering
        \includegraphics[width=0.44\textwidth]{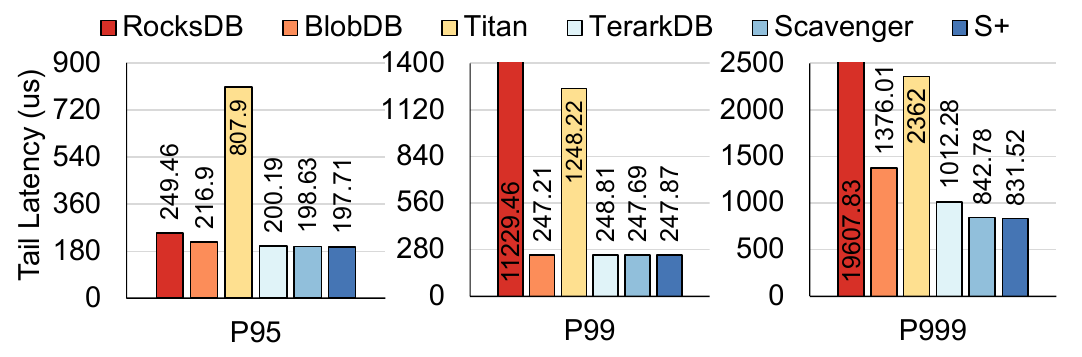}
        \caption{Tail latencies of update under Mixed-8K without space limits.}
        \label{fig:test99-tail-lat}
        \vspace{-0.5cm}
\end{figure}

\begin{figure}[tbp]
        \setlength{\abovecaptionskip}{0pt}
        \setlength{\belowcaptionskip}{-0.6cm}
	\centering
        \includegraphics[width=0.44\textwidth]{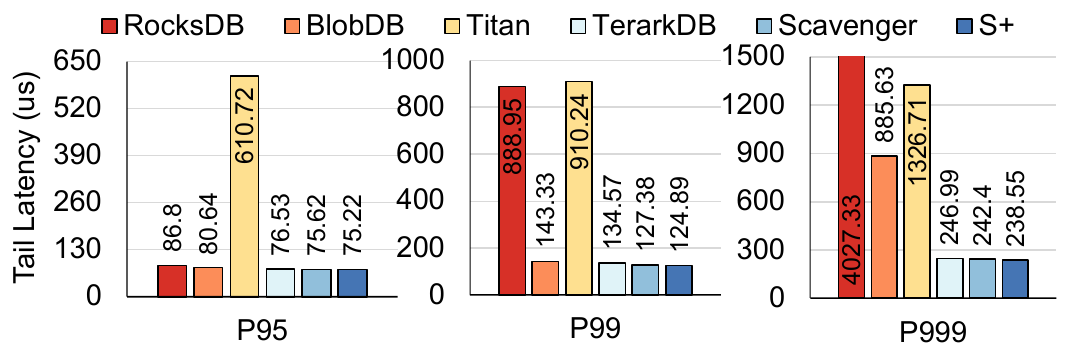}
        \caption{Tail latencies of update under Pareto-1K without space limits.}
        \label{fig:test98-tail-lat}
        \vspace{-0.5cm}
\end{figure}

\textbf{Mixed-8K workload}. As depicted in Figure~\subref*{fig:test1-micro-mixed-iops}, Scavenger+ (\textbf{\textit{S+}}) significantly enhances update performance and outperforms RocksDB, BlobDB, Titan, and TerarkDB by 3.2x, 2.9x, 3.1x, and 2.6x, respectively. Compared to Scavenger, Scavenger+ improves update performance by 1.2x due to adaptive GC readahead and dynamic GC scheduling. It also demonstrates 1.3x better read performance than RocksDB, due to prioritized caching of index data that accelerates large-value access. For insertion and scanning, Scavenger+ performs comparably to TerarkDB. For insertion, all KV-separated LSM-trees outperform RocksDB by 4-5x. In scanning, frequent compactions in RocksDB improve value order, outperforming KV-separated LSM-trees. Notably, Scavenger+ and TerarkDB outperform others due to TerarkDB's efficient GC operations that enhance data order. Since only update workloads affect space amplification, Scavenger+ achieves the most substantial performance gains in update operations with limited space while preserving TerarkDB's advantages in other operations.

Given Scavenger+'s significant enhancement in update performance, we evaluated its read and write I/O under the update workload, as shown in Figure \subref*{fig:test1-micro-ndb-io}. Compared to other systems, the read I/O of Scavenger+ constitutes only 73\% of that of BlobDB, 66\% of Titan, and 52\% of TerarkDB. For write I/O, Scavenger+ achieves a maximum reduction of 28\%. Additionally, we examined the I/O associated with GC operations in KV-separated LSM-trees. This analysis shows that the reduction in disk read and write I/O for Scavenger+ is mainly due to minimized I/O during GC operations. Notably, the read/write I/O for Scavenger+ is slightly higher than Scavenger's, due to increased GC concurrency and more GC operations, but this does not negatively affect performance.

\textbf{Pareto-1K workload}. As depicted in Figure~\subref*{fig:test1-micro-pareto-iops}, Scavenger+ significantly boosts update performance, showing improvements ranging from 1.3x to 5.9x. Scavenger's update performance is similar to RocksDB since small-value workloads reduce the benefits of KV separation. However, with adaptive GC readahead, Scavenger+ mitigates the impact of small I/Os and outperforms RocksDB. Scavenger+'s insertion and read performance align with the Mixed-8K workload findings, where it also showed superior performance. In scanning, Scavenger+ continues to outperform other KV-separated LSM-trees but does not match RocksDB's performance, as it compromises insert performance to improve data ordering.

\textbf{Without space limits}. We evaluated the performance and space amplification of Mixed-8K and Pareto-1K workloads without space constraints. Since only write operations affect space usage, our focus was on performance and space amplification during updates. As shown in Figure~\ref{fig:test2-micro-nolimit}, Scavenger+ achieves the highest throughput in both workloads while maintaining lower space amplification (1.71 and 1.60) than other KV-separated LSM-trees. Additionally, we assessed the tail latency of update operations. Figures~\ref{fig:test99-tail-lat} and \ref{fig:test98-tail-lat} demonstrate that Scavenger+ achieves the lowest tail latency, sustaining high throughput with minimal latency impact. We further analyzed P99 latency fluctuations over 1500 seconds in Section D of the supplementary file, where Scavenger+ achieves the best performance stability. Compared to Scavenger, the adaptive readahead and dynamic scheduling in Scavenger+ enable faster GC execution within allocated resources, reducing space amplification without compromising performance.

\subsection{YCSB Evaluation}
We evaluated the performance of Scavenger+ by employing YCSB workloads \cite{cooper2010benchmarking}, focusing specifically on throughput after extensive updates. For each YCSB workload, we followed this procedure: initializing with 100GB of uniformly distributed data, then applying 300GB of updates to ensure GC activation across all KV-separated LSM-trees. Subsequently, we executed YCSB workloads A-F on the updated dataset, maintaining the same configurations as previously described.

\begin{figure*}[tbp]
        \setlength{\abovecaptionskip}{5pt}
        \setlength{\belowcaptionskip}{-0.9cm}
	\centering
	\subfloat[Throughput under Mixed-8K]{
            \includegraphics[width=.40\textwidth]{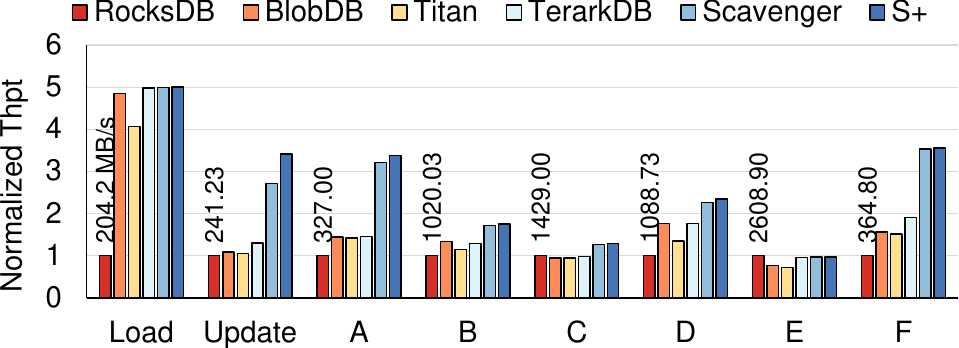}
            \label{fig:test3-ycsb-mixed}
        }
        \subfloat[Throughput under Pareto-1K]{
            \includegraphics[width=.40\textwidth]{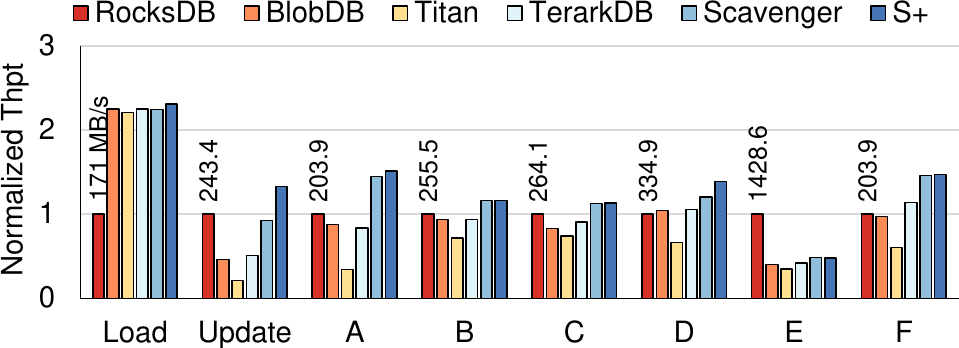}
            \label{fig:test3-ycsb-pareto}
        }
	\caption{YCSB under Mixed-8K and Pareto-1K with 1.5x space limit.}
        \vspace{-0.9cm}
\end{figure*}

\begin{figure}[t]
        \setlength{\abovecaptionskip}{5pt}
        \setlength{\belowcaptionskip}{-0.6cm}
	\centering
	\subfloat[Throughput]{
            \includegraphics[width=.46\columnwidth]{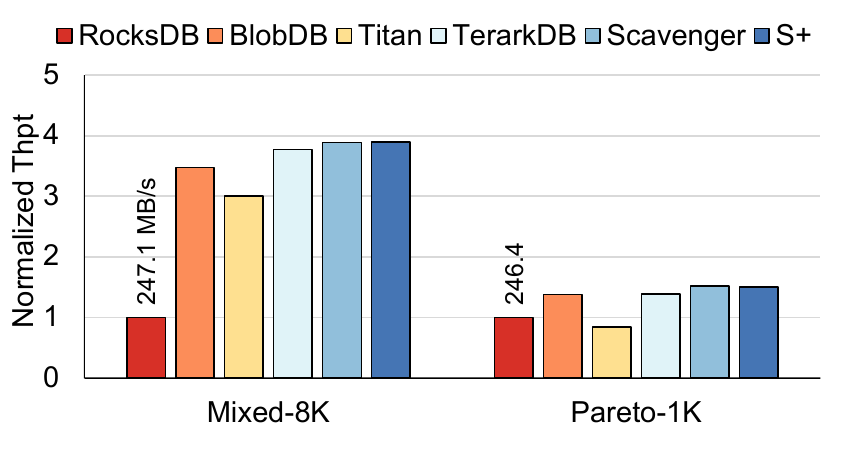}
            \label{fig:test4-ycsb-nolimit-a}
        }
        \subfloat[Space amplification]{
            \includegraphics[width=.46\columnwidth]{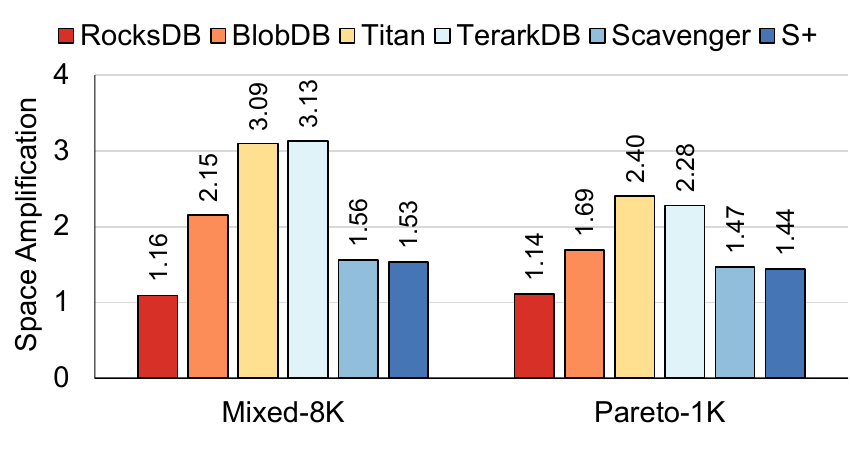}
            \label{fig:test4-ycsb-nolimit-space}
        }
	\caption{YCSB-A without space limits.}
        \label{test4-ycsb-nolimit}
        \vspace{-0.5cm}
\end{figure}

\begin{figure}[tbp]
        \setlength{\abovecaptionskip}{5pt}
        \setlength{\belowcaptionskip}{-0.6cm}
	\centering
	\subfloat[Compaction and GC Features]{
            \includegraphics[width=.55\columnwidth]{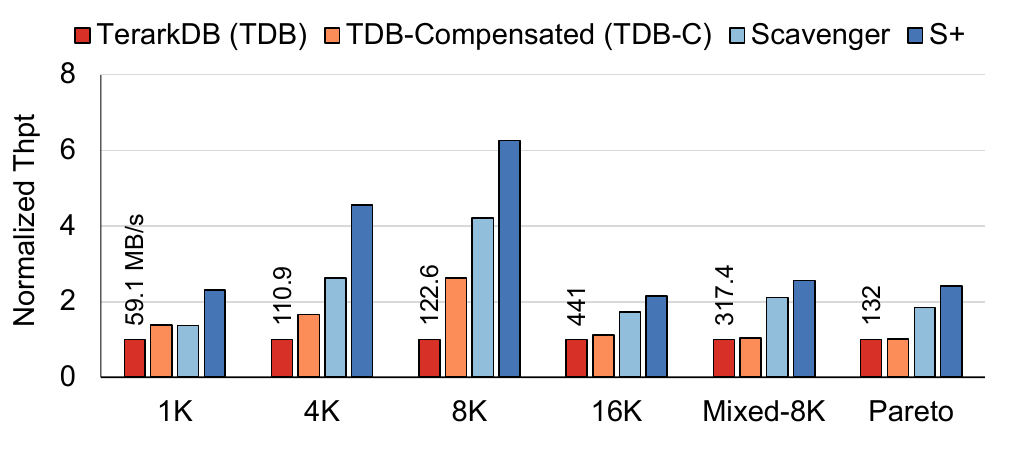}
            \label{fig:test5-features-all}
        }
        \subfloat[GC Features (RWL)]{
            \includegraphics[width=.38\columnwidth]{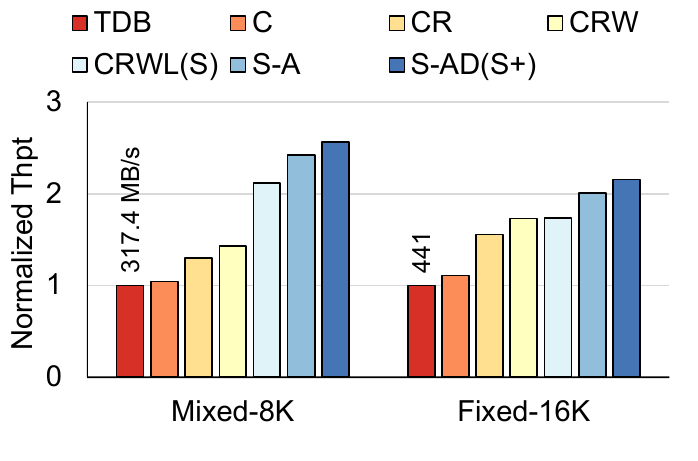}
            \label{fig:test5-features-gc}
        }
	\caption{Write throughput of features under 1.5x space limit.}
         \label{fig:test5-features}
         \vspace{-0.5cm}
\end{figure}

\textbf{Mixed-8K workload}. Initially, we evaluated the performance under Mixed-8K workloads with a 1.5x space constraint, as shown in Figure~\subref*{fig:test3-ycsb-mixed}. For write-intensive workloads, Scavenger+ significantly improves performance. Specifically, compared to other KV-separated LSM-trees, Scavenger+ showed performance improvements ranging from approximately 1.1x to 2.4x in the YCSB-A mixed-read-write (1:1) workload and 1.0x to 2.3x in the YCSB-F workload. Compared to Scavenger, Scavenger+ achieves the greatest improvement in update while also improving performance in mixed-read-write workloads. In read-intensive workloads, Scavenger+ exhibits performance comparable to Scavenger and surpasses all other alternatives. Under the YCSB-E workload, both Scavenger/Scavenger+ and TerarkDB perform on par with RocksDB, surpassing other KV-separated LSM-trees.

We conducted evaluations with the Pareto-1K workload. As Figure \ref{fig:test3-ycsb-pareto} shows, Scavenger+ shows superior performance across all workloads except YCSB-E. Consistent with the findings under Mixed-8K, Scavenger+ demonstrates enhanced performance in write-intensive workloads, including YCSB-A and YCSB-F. For YCSB-E, RocksDB improves data ordering by reducing write performance. It is effective because smaller KV pairs under the Pareto-1K are strongly influenced by data order, allowing it to outperform all KV-separated LSM-trees. Scavenger+ outperforms other KV-separated LSM-trees due to efficient GC operations that enhance data ordering in vSSTs.

\textbf{Without space limits}. In this section, we focus specifically on write-intensive workloads because of their strong association with space amplification, using YCSB-A as a representative example. Figure \ref{test4-ycsb-nolimit} shows the results for both Mixed-8K and Pareto-1K workloads. The insights obtained here align with those from the microbenchmark tests, where Scavenger and Scavenger+ demonstrate superior foreground performance and significantly lower space amplification compared to other KV-separated LSM-trees. Notably, the performance and space amplification of both Scavenger+ and Scavenger closely align in the unrestricted space setting of YCSB-A. This is because GC pressure in a mixed read-write workload scenario is less intense than in a purely write-intensive workload, as Scavenger is well-equipped to handle such conditions.
\begin{figure}[tp]
        \setlength{\abovecaptionskip}{5pt}
        \setlength{\belowcaptionskip}{-0.6cm}
	\centering
	\subfloat[Compaction and GC Features]{
            \includegraphics[width=.53\columnwidth]{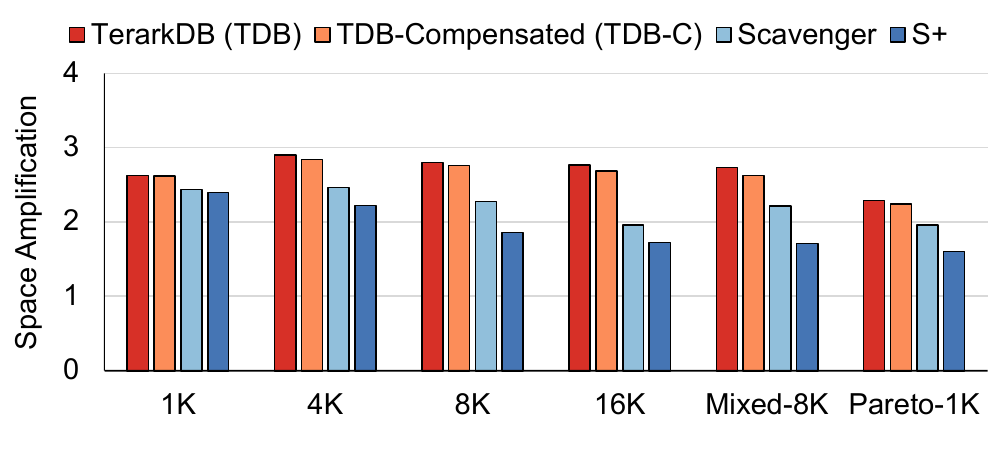}
            \label{fig:test6-features-all}
        }
        \subfloat[GC Features (RWL)]{
            \includegraphics[width=.40\columnwidth]{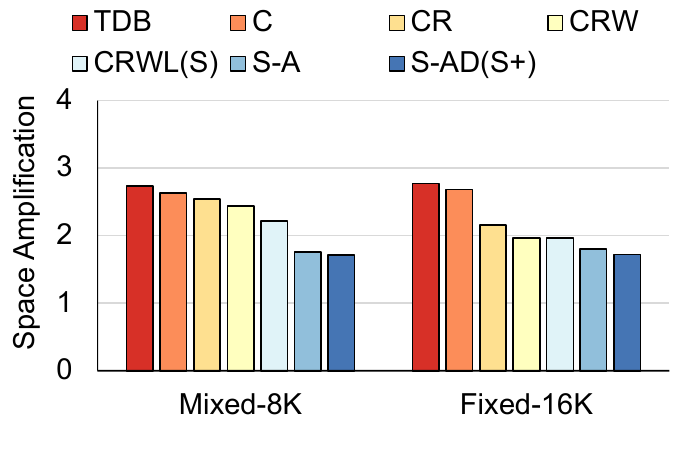}
            \label{fig:test6-features-gc}
        }
	\caption{Space amplification of features without space limits.}
         \label{fig:test6-features-1}
         \vspace{-0.5cm}
\end{figure}

\begin{figure}[tbp]
        \setlength{\abovecaptionskip}{5pt}
        \setlength{\belowcaptionskip}{-0.6cm}
	\centering
	\subfloat[Space amplification of index]{
            \includegraphics[width=.47\columnwidth]{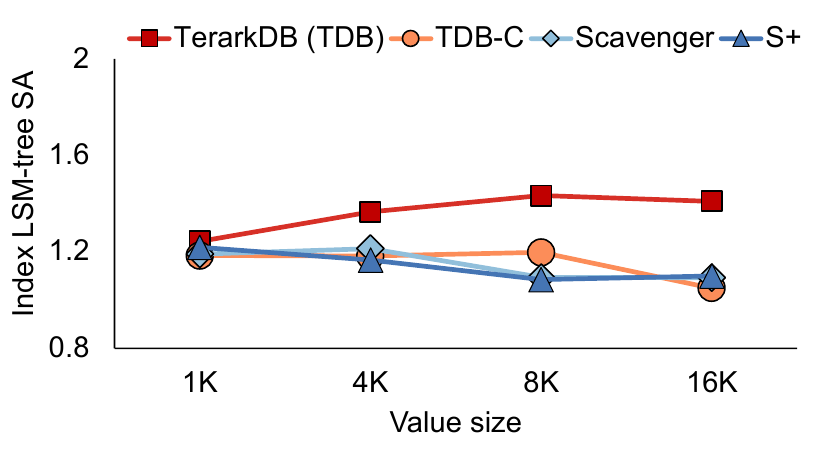}
            \label{fig:test6-features-index-sa}
        }
        \subfloat[Exposed garbage of value]{
            \includegraphics[width=.47\columnwidth]{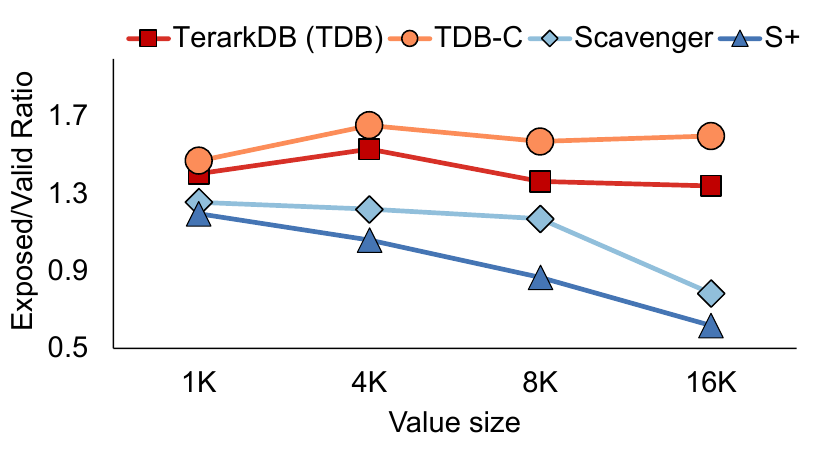}
            \label{fig:test6-features-exposed-sa}
        }
	\caption{Sources of space amplification without space limits.}
        \label{test6-features}
        \vspace{-0.5cm}
\end{figure}

\subsection{Scavenger+ Features}
We assessed the impact of each optimization technique implemented in Scavenger+. Evaluations focused on quantifying the performance benefits and storage overhead of each technique by examining write performance with limited storage costs and space amplification in scenarios without space limits. Specifically, we tested TerarkDB without optimizations (\textbf{TDB}), the space-aware compaction strategy (\textbf{TDB-C}), the I/O-efficient GC as proposed by \textbf{Scavenger}, and the adaptive readahead and dynamic GC scheduling features introduced by Scavenger+ (\textbf{S+}). Furthermore, we assessed Scavenger+'s adaptability by examining its performance and space amplification across diverse workloads with varying value sizes. Additionally, we explored the impact of optimizing various I/O-efficient GC features, encompassing lazy read (\textbf{CR}), hotspot-aware write (\textbf{CRW}), GC-Lookup optimized by separating records and indexes (\textbf{CRWL, S}), adaptive readahead (\textbf{S-A}), and dynamic GC scheduling (\textbf{S-AD, S+}).

\textbf{Write performance with limited space}. Figure \subref*{fig:test5-features-all} indicates that the space-aware compaction strategy improves update performance by 1.4 to 2.6 times for fixed-length workloads. For variable-length workloads, the benefits of the space-aware compaction strategy are less pronounced as the index LSM-tree accommodates many key-value pairs with larger average value sizes, leading to a significant increase in both file size and levels. Concurrently, I/O-efficient GC is effective in most scenarios, especially for workloads with large values and variable lengths. Figure \subref*{fig:test5-features-gc} illustrates how lazy read (\textbf{R}) significantly enhances write performance for large-value workloads, especially in fixed-length scenarios, while GC-Lookup optimization (\textbf{L}) excels in managing variable-length workloads. Hotspot-aware write (\textbf{W}) can improve write performance by 10\% for workloads with hotspots. Adaptive readahead (\textbf{S-A}) further improves write performance even for small-value workloads by reducing IOPS through larger I/O granularity. Dynamic GC scheduling (\textbf{S-AD}) improves write performance by optimizing thread resource utilization.

\textbf{Space amplification without space limits}. We evaluated write performance and space amplification without space limits. Scavenger+, Scavenger, TerarkDB, and TDB-C exhibited comparable write performance, leading us to focus on space amplification. Figure~\subref*{fig:test6-features-all} illustrates that space-aware compaction can reduce space amplification by up to 4\%, as this targets the index LSM-tree exclusively. These findings highlight the importance of I/O-efficient GC, which can cut space amplification by as much as 38\%. Figure~\ref{test6-features} provides a detailed root cause analysis. Space-aware compaction reduces the index LSM-tree's space amplification ($S_{index}$) to approximately 1.1, but it increases the ratio of exposed garbage in value data ($G_e/D$). Without GC optimization, the system cannot quickly reclaim the substantial volume of generated garbage, resulting in a minimal reduction in space amplification. Moreover, we examined the impact of different GC optimization techniques, as shown in Figure~\subref*{fig:test6-features-gc}. Similar to the performance results, GC-Lookup optimization (\textbf{L}) shows the greatest impact on variable-length workloads, while lazy read optimization (\textbf{R}) is the most impactful for fixed-length workloads. Write optimization (\textbf{W}) benefits all tested workload types. Adaptive readahead (\textbf{S-A}) and dynamic GC scheduling (\textbf{S-AD}) both contribute to reduced space amplification.

\section{Related Work}
Recently, researchers have been actively working on improving the write performance of LSM-tree-based KVS. Here, we present two main directions and discuss the differences between their proposals and our solution.

\textbf{Compaction Optimization}. Write performance degradation in LSM-trees primarily arises from write amplification induced by strict key ordering during compaction, motivating researchers to explore order relaxation strategies. PebblesDB \cite{raju2017pebblesdb} employs a fragmented LSM-tree (FLSM) to alleviate single-level orderliness. Other schemes, such as UniKV \cite{zhang2020unikv}, WipDB \cite{zhao2021wipdb}, and REMIX \cite{zhong2021remix}, leverage partitioning to minimize compaction overhead and do not guarantee order in the single partition strictly, while BlockDB \cite{wang2022reducing} optimizes compaction granularity and strategy to mitigate write amplification. L2SM \cite{huang2021less} and LSbM \cite{teng2017lsbm} utilize buffers to attain an effect similar to tiering compaction. However, these designs may be affected by read and space amplification issues commonly associated with tiering compaction. Sarkar et al. \cite{sarkar2021constructing, sarkar2022compactionary} introduced a design space for compaction in vanilla LSM-trees but did not address KV-separated LSM-tree-specific challenges. While some strategies help reduce the space amplification of vanilla LSM-trees, their applicability to KV-separated LSM-trees is limited due to the lack of visibility into separated value sizes during compaction.

\textbf{Key-Value Separation}. Key-value separation has attracted wide attention in both academia and industry due to its simplicity and effectiveness. The fundamental principle is to reduce the data involved in compaction, thus mitigating write amplification during compaction. WiscKey \cite{lu2017wisckey} and HashKV \cite{chan2018hashkv} pioneered KV separation but paid less attention to query performance, particularly for range queries. Bourbon extends WiscKey with learning indexes \cite{dai2020wisckey} to improve read efficiency; however, it retains WiscKey's unordered storage layout and is less effective under dynamic workloads. TerarkDB \cite{terarkdb}, Titan \cite{titan}, DiffKV \cite{li2021differentiated}, and BlobDB \cite{rocksdb} enhanced range queries by utilizing structures similar to the original SST of RocksDB. Nevertheless, these systems may still face performance variability stemming from GC overhead. For instance, Titan may experience contention between foreground writes and write-back GC, while the compaction-triggered GC in DiffKV and BlobDB can couple GC with compaction, potentially limiting reclamation efficiency. Most existing KV-separated LSM-trees do not explicitly address space amplification or consider its relation to the shrinking size of the index LSM-tree. Contrarily, Scavenger+ identifies the two primary causes of space amplification in KV-separated LSM-trees and accelerates space reclamation by reducing GC I/O overhead while also optimizing compaction to enhance GC efficiency. Consequently, Scavenger+ achieves a better balance between performance and space amplification.

\section{Conclusion}
In this paper, we present Scavenger+, a key-value store based on a KV-separated LSM-tree. We analyze the root causes of space amplification in KV-separated LSM-trees and introduce an I/O-efficient garbage collection scheme to reduce I/O overhead. Additionally, we integrate a space-aware, size-compensated compaction strategy to mitigate space amplification in index LSM-trees. By reducing I/O overhead and optimizing task scheduling for resource efficiency, Scavenger+ rapidly reclaims space while preserving foreground performance, resulting in superior space-time trade-offs.

\bibliographystyle{IEEEtran}
\bibliography{IEEEabrv, tc}

\vfill

\end{document}